\documentclass[letterpaper]{article}

\usepackage[english]{babel}
\usepackage{siunitx}
\usepackage[margin=1in]{geometry}

\usepackage[numbers,comma,sort&compress]{natbib}
\makeatletter
\providecommand{\doi}[1]{%
  \begingroup
    \let\bibinfo\@secondoftwo
    \urlstyle{rm}%
    \href{http://dx.doi.org/#1}{%
      doi:\discretionary{}{}{}%
      \nolinkurl{#1}%
    }%
  \endgroup
}
\makeatother

\usepackage{amsmath}
\usepackage{amssymb}
\usepackage{bm}
\usepackage{mathrsfs}
\DeclareMathOperator{\NN}{NN}
\newcommand{\MSE}[0]{\mathrm{MSE}}

\usepackage{graphicx}
\graphicspath{{./figs/}}
\usepackage{float}

\usepackage{fancyhdr}
\usepackage{totpages}
\fancypagestyle{firstpage}{%
  \fancyhf{}%
  \fancyfoot[C]{\thepage\ of \ref*{TotPages}}%
}

\usepackage{hyperref}
\hypersetup{
    colorlinks=true,
    linkcolor=blue,
    citecolor=blue,
    filecolor=magenta,      
    urlcolor=cyan,
    breaklinks=true
}

\allowdisplaybreaks

\title{\textbf{Physics-informed neural networks for understanding shear migration of particles in viscous flow}}
\author{Daihui Lu and Ivan C.\ Christov\footnote{Corresponding author. \href{mailto:christov@purdue.edu}{\texttt{christov@purdue.edu}}; \url{http://tmnt-lab.org}.}\\[2mm]
{\small\textit{School of Mechanical Engineering, Purdue University, West Lafayette, Indiana 47907, USA}}}

\begin{document}

\maketitle 
\thispagestyle{firstpage}

\begin{abstract}
    We harness the physics-informed neural network (PINN) approach to extend the utility of phenomenological models for particle migration in shear flow. Specifically, we propose to constrain the neural network training via a model for the physics of shear-induced particle migration in suspensions. Then, we train the PINN against experimental data from the literature, showing that this approach provides both better fidelity to the experiments, and a novel understanding of the relative roles of the hypothesized migration fluxes. We first verify the PINN approach for solving the inverse problem of radial particle migration in a non-Brownian suspension in an annular Couette flow. In this classical case, the PINN yields the same value (as reported in the literature) for the ratio of the two parameters of the empirical model. Next, we apply the PINN approach to analyze experiments on particle migration in both non-Brownian and Brownian suspensions in Poiseuille slot flow, for which a definitive calibration of the phenomenological migration model has been lacking. Using the PINN approach, we identify the unknown/empirical parameters in the physical model through the inverse solver capability of PINNs. Specifically, the values are significantly different from those for the Couette cell, highlighting an inconsistency in the literature that uses the latter value for Poiseuille flow. Importantly, the PINN results also show that the inferred values of the empirical model's parameters vary with the shear P\'eclet number and the particle bulk volume fraction of the suspension, instead of being constant as assumed in some previous literature.
    
\end{abstract}

\section{Introduction}

\citet{pabga92} proposed an insightful phenomenological model for the shear-induced migration of particles in a low-Reynolds-number flow \citep{la87}. Specifically, they posited that the distribution of particles, accounted for by the volume fraction $\phi(\bm{x},t)$ of the fluid--particle suspension, obeys a conservation law \cite{pabga92,MS95}:
\begin{equation}
    \frac{D \phi}{D t} = -\bm{\nabla} \bm{\cdot} \bm{J}.
    \label{eq:continuity_suspension}
\end{equation}
In Eq.~\eqref{eq:continuity_suspension}, the material derivative of the left-hand side represents the unsteady transport of particles by a flow $\bm{u}$, while $\bm{J}$ on the right-hand side represents a spatial flux arising from the hydrodynamic interactions of particles. Consequently, Eq.~\eqref{eq:continuity_suspension} is often referred to as a \emph{diffusive-flux model} in the literature. \citet{pabga92} decomposed the flux as $\bm{J}= \bm{N}_c + \bm{N}_{\eta} + \bm{N}_B$, where they posited that $\bm{N}_c$, $\bm{N}_{\eta}$ and $\bm{N}_B$ represent the particle fluxes due to the variations in the particle collision frequency, the spatial variations of the viscosity of the suspension (due to the nonuniform particle distribution \cite{GMA1980,la87}), and the spatial variations of the concentration (responsible for Brownian diffusion, by Fick's law), respectively. 
For unidirectional flows, ${D \phi}/{D t} = {\partial \phi}/{\partial t}$ \cite{MS95}. 

Specifically, \citet{pabga92} proposed the following ``constitutive laws'' for the diffusive fluxes:
\begin{subequations}\begin{align}
    \bm{N}_c &= -K_ca_p^2 \left(\phi^2\bm{\nabla} \dot{\gamma} + \phi \dot{\gamma} \bm{\nabla}\phi\right) = -K_c a_p^2 \phi \bm{\nabla}(\phi\dot{\gamma}),\\
    \bm{N}_{\eta} &= -K_{\eta} \dot{\gamma}\phi^2 \left(\frac{a_p^2}{\eta}\right) \frac{d\eta}{d\phi} \bm{\nabla} \phi = -K_\eta a_p^2 \phi^2\dot{\gamma} \bm{\nabla}(\ln\eta),\\
    \bm{N}_B &= -D\bm{\nabla}\phi,
\end{align}\label{eq:Ns_def}\end{subequations}
where $K_c$ and $K_{\eta}$ are \textit{a priori} unknown constants of order unity, which are found from experimental data (by fitting/calibration). Therefore, Eq.~\eqref{eq:continuity_suspension}, with the fluxes given in Eqs.~\eqref{eq:Ns_def}, becomes a \emph{parameterized partial differential equation} (PDE).
Here, $a_p$ is a particle's radius, $D$ is its Brownian diffusivity (in principle, known from the Stokes--Einstein relation $D=k_B T/{6\pi \eta_s a_p}$ with $T$ being temperature and $k_B$ being Boltzmann's constant), $\eta$ is the non-constant dynamic viscosity of the suspension, which may depend on many parameters \citep{SP05}, $\eta_s$ is the carrier Newtonian fluid's viscosity, and $\dot{\gamma}(\bm{x},t)$ is the non-uniform shear rate in the flow. For a general flow field $\bm{u}=\bm{u}(\bm{x},t)$, the shear rate is evaluated as the magnitude of the rate-of-strain tensor $\bm{E}=\frac{1}{2}(\bm{\nabla}\bm{u} + \bm{\nabla}\bm{u}^\top)$, i.e., $\dot{\gamma}=\sqrt{2\bm{E}:\bm{E}}$.

At steady state, $\partial(\,\cdot\,)/\partial t = 0$. Then, Eq.~\eqref{eq:continuity_suspension} can be integrated once in space over some domain $\mathcal{V}$, with the constant of integration set to zero by imposing a no-flux condition on the domain's boundary $\partial\mathcal{V}$. Therefore, the resulting governing physics equation at steady-state is 
\begin{equation}
    \bm{N}_c + \bm{N}_{\eta} + \bm{N}_B = \bm{0},
    \label{eq:steady_susp}
\end{equation}
which is a transport equation expressing the conservation of particles. Equation~\eqref{eq:steady_susp} also implies that if the initial particle volume fraction is such that $\int_\mathcal{A}\phi(\bm{x},0) \,d\bm{x} = \phi_b = const.$ across any cross-section $\mathcal{A}$, then  $\int_\mathcal{A} \phi(\bm{x},t) \,d\bm{x} = \phi_b$ for any $t>0$. Additionally, the flow field $\bm{u}$ obeys the low-Reynolds-number (inertialess) flow momentum equation \citep{pabga92}, which takes the form (using a more standard \citep{Panton} sign convention and definition of $\bm{E}$ as above):
\begin{equation}
    \bm{\nabla}\bm{\cdot}\bm{\tau} = \bm{\nabla}p, \qquad \bm{\tau} = 2\eta\bm{E},\qquad \eta =\eta_s\eta_r(\phi),
\label{eq:mom_steady}
\end{equation}
where $\eta_r$ (dimensionless) is the contribution from the suspension to be introduced below, and $p(\bm{x})$ is the hydrodynamic pressure. We will consider only neutrally buoyant suspensions, and so body forces are neglected in Eq.~\eqref{eq:mom_steady}. The velocity field is additionally incompressible,  $\bm{\nabla}\bm{\cdot}\bm{u}=0$, but this relation is automatically satisfied by the unidirectional flows considered herein \citep{Panton}, so it is not a physical constraint that we need to enforce explicitly.

The fluid mechanics of particulate suspensions remains a frontier problem \citep{M20}, and the diffusive-flux model of \citet{pabga92} is not without its criticisms  \cite{MS95}. Nevertheless, although much more sophisticated models of suspensions exist \citep{vsb10,GuaMor}, including the suspension balance model \citep{NB94,MB1999,FMBIMG02,MM2006,L09,Nott2011}, the two-fluid model \citep{MVP17,MNC19}, even direct numerical simulation \citep{M17}, Eqs.~\eqref{eq:continuity_suspension}--\eqref{eq:Ns_def} remain a popular model through which to study shear-induced particle migration in suspensions \citep{KM20,FSM20,H21}.  

``Disentagling'' the individual effects of shear-induced fluxes in Eq.~\eqref{eq:steady_susp} has been of particular interest in the suspensions literature \citep{MLBM05}. (Note that although \citet{MLBM05} also included a fourth, ``curvature-induced'' flux in Eq.~\eqref{eq:steady_susp}, its origin has been disputed by \citet{BB06}.) Machine learning is a viable approach toward processing experimental data to disentangle the relative strengths of the fluxes in Eq.~\eqref{eq:steady_susp}. To this end, in Section~\ref{sec:PINN_algo}, we apply the  physics-informed neural network (PINN) approach of \citet{RPK19} towards understanding particle migration in shear flow. Specifically, we propose to constrain  the neural network using the model given by Eqs.~\eqref{eq:steady_susp} and \eqref{eq:mom_steady}. In Section~\ref{sec:couette}, we validate this approach on the classical Couette cell experiments (and modeling) of \citet{pabga92}. Then, in Section~\ref{sec:Poiseuille}, we apply the PINN approach to the more challenging case of Poiseuille channel flow. In doing so, we re-interpret experiments on pressure-driven flows of both non-Brownian (Section~\ref{sec:Poiseuille_NB}) and Brownian (Section~\ref{sec:Brownian}) suspensions, uncovering new aspects of the shear-induced migration model. Finally, conclusions are drawn in Section~\ref{sec:conclusion}. Additional data preprocessing and verification details (regarding the PINN calculations) are provided in Appendices~\ref{sec:pre-proc} and \ref{sec:choice_layers}, respectively.

\section{PINN algorithm description and implementation}
\label{sec:PINN_algo}

In recent years, with the explosive growth of available data, computing modalities, and requisite hardware resources, deep learning algorithms have been applied to a range of problems arising from computer science, physics and engineering fields \citep{BK19_book}, including in the field of fluid mechanics \citep{BNK20}. 
\emph{Physics-informed machine learning} has emerged as an approach that ``integrates seamlessly data and mathematical physics models, even in partially understood ... contexts'' \citep{KKLPWY21}. Specifically, \citet{RPK19} developed a computational approach to couple machine learning with some underlying physics (human reasoning), which they termed \emph{physics-informed neural networks} (PINNs). PINNs are a deep learning framework for solving problems involving PDEs by embedding (in a suitable sense) the physics into the neural network. Due to their versatility, PINNs have been applied to solve forward and inverse problems in fluid mechanics \citep{YZWX19,MJK19,HT20,JCLK20,RHPT20,MYK20}, solid mechanics \citep{HRMGJ21,LBZ21,HWM22}, and heat transfer \citep{CWWPK21}, amongst many other applications. PINNs are appealing due to their standardized implementation. They use automatic differentiation \cite{BPRS18} techniques to discretize the differential operators needed for the back-propagation problem, as well as the PDEs representing the physics \citep{RPK19,MK20}. This approach makes PINNs mesh-free and, thus, easy to use for evaluating the PDE residual even from sparse experimental data sets (`observations').  Importantly, PINNs can determine unknown parameters in the physics embedded therein, even from incomplete (or partial) data sets, making PINNs useful for reduced-order model calibration.

\begin{figure}[t]
  \centering
  \includegraphics[width=0.85\textwidth]{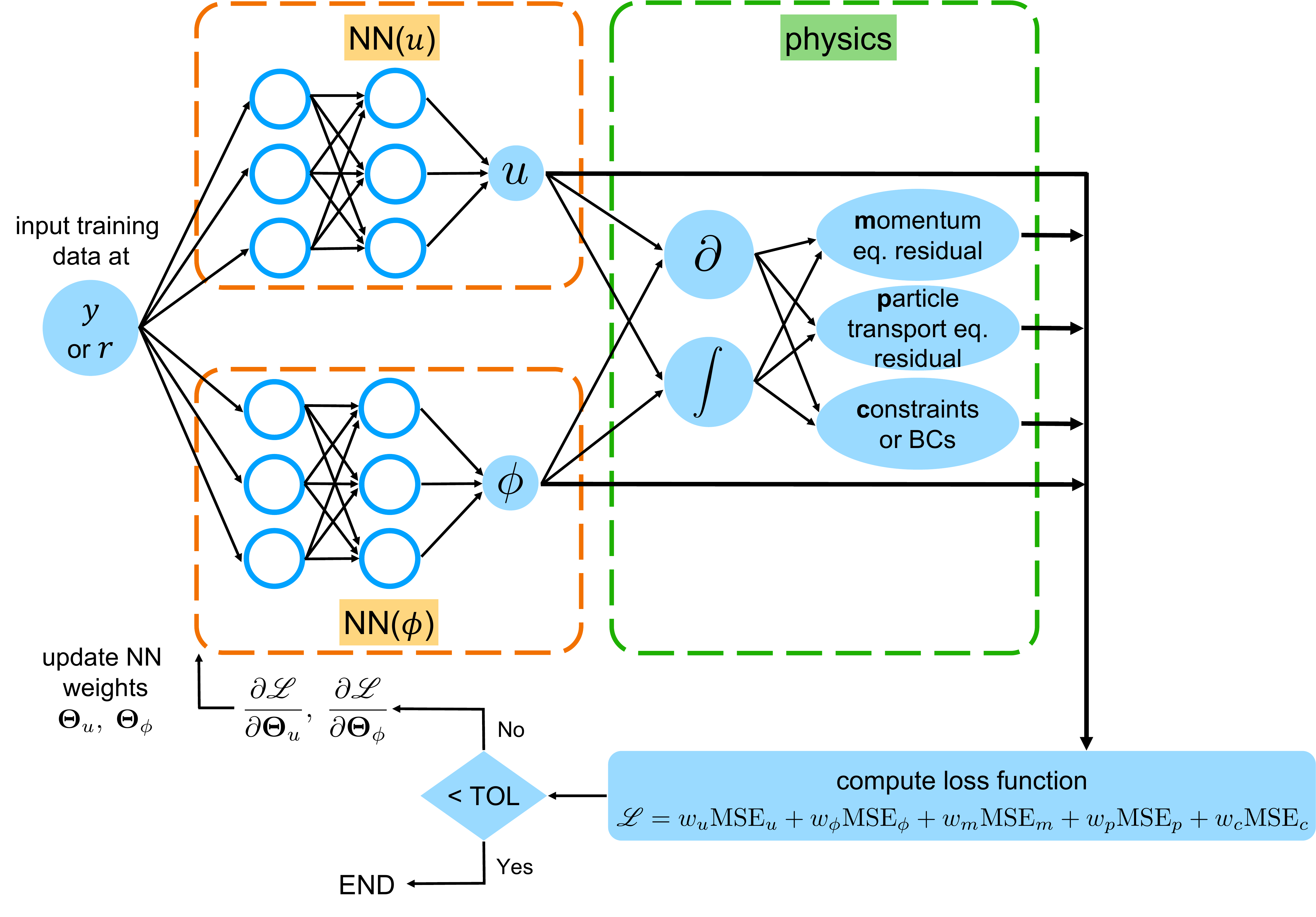}
\caption{Architecture of the proposed PINN for solving the inverse problem of reduced-model determination for particle migration in suspensions. The loss $\mathscr{L}$ is formulated (as in Eq.~\eqref{eq:loss}) in terms of root-mean-squared errors between predictions and observations ($\MSE_u$, $\MSE_{\phi}$), and errors in satisfaction (residuals) of the underlying PDEs from the physics ($\MSE_m$, $\MSE_p$), as well as boundary conditions and/or additional constraints ($\MSE_{c}$).}
\label{fig:pinn_diagram}
\end{figure}

In this paper, we use a PINN to solve the inverse problem of reduced-model determination for particle migration in suspensions. Given measurements of a velocity field $\bm{u}$ and a particle volume fraction profile $\phi$, we seek to learn the unknown parameters $K_c$ and $K_{\eta}$ in the fluxes given in Eqs.~\eqref{eq:Ns_def}. 
The governing physics equations are embedded into the PINN as shown in Fig.~\ref{fig:pinn_diagram}.
We use two independent neural networks, $\NN(u)$ and $\NN(\phi)$, to approximate the velocity distribution and particle distribution, respectively. Both NNs are fully-connected and feed-forward, with multiple hidden layers each. 

Suppose that the measured data is available on $N$ (possibly random) sample points. The residuals of the fluid's conservation of momentum equation~\eqref{eq:mom_steady} and the suspension's particle transport equation~\eqref{eq:steady_susp} (suitably simplified for some given flow conditions and domain geometry) are evaluated from the approximated values of $\bm{u}$ and $\phi$ at these $N$ collocation points. Then, combining the error between predictions and observations with the error in satisfying the physics from the residuals, along with any constraints, we formulate a loss function as:
\begin{equation}
    \mathscr{L} = \underbrace{w_u \MSE_u +w_{\phi} \MSE_{\phi}}_{\text{training data}} + \underbrace{w_m \MSE_m + w_p \MSE_p}_{\text{physics}} +  \underbrace{w_{c} \MSE_{c}}_{\text{constraints}},
    \label{eq:loss}
\end{equation}
where, for example, 
\begin{subequations}\begin{align}
\MSE_u &= \frac{1}{N}\sum_{i = 1}^{N} \|\bm{u}(i) - \bm{u}_\mathrm{train}(i)\|^2, \\
\MSE_{\phi}  &= \frac{1}{N}\sum_{i = 1}^{N} |\phi(i) - \phi_\mathrm{train}(i)|^2 , \\
\MSE_m &= \frac{1}{N}\sum_{i = 1}^{N}\left\|\bm{\nabla}\bm{\cdot}\bm{\tau}(i) - \bm{\nabla}p(i)\right\|^2,\\
\MSE_p &= \frac{1}{N}\sum_{i = 1}^{N}\left\|\bm{N}_c(i) + \bm{N}_{\eta}(i) + \bm{N}_B(i)\right\|^2.
\end{align}
\end{subequations}

The $\MSE$ terms in $\mathscr{L}$ represent various ``mean squared errors.'' The notation ``$(i)$'' denotes the value of the quantity at the $i$th data point in the set of $N$ observations. For clarity, we omit this explicit notation below without fear of confusion. The first two terms of $\mathscr{L}$ correspond to the errors between the predicted and the input velocity and particle distribution training data, respectively. Then, the following two terms of $\mathscr{L}$  correspond to the error in satisfaction of the physics, \textit{i.e.}, the suspension's momentum equation~\eqref{eq:mom_steady} and particle transport equation~\eqref{eq:steady_susp}, respectively. The last term of $\mathscr{L}$  represents error committed in satisfaction of ``constraints.'' The constraints can involve, \textit{e.g.}, boundary conditions, integral constraints, or any other mathematical statement not captured in the ``physics'' term, which is typically used to denote only the satisfaction of governing (partial) differential equations. The coefficients $w_j$ where $j \in \{ u, \phi, p, m, c \}$, represent weights of the corresponding loss terms. Although the relative values of the weights of terms in the loss function may influence the ability to train the NN \cite{WYP20}, here we generally take them to be equal.

Figure~\ref{fig:pinn_diagram} shows the architecture of the PINN for the case in which only a component $u$ of the velocity vector $\bm{u}$ has been measured. Initially, a randomly selected set of network weight vectors $\bm{\Theta}_u^{(0)}$ and $\bm{\Theta}_\phi^{(0)}$ are used to construct $\mathrm{NN}(u)$ and $\mathrm{NN}(\phi)$, respectively. Then, we feed training data into $\mathrm{NN}(u)$ and $\mathrm{NN}(\phi)$ to obtain predictions on $u$ and $\phi$. We calculate the derivatives of $u$ and $\phi$ needed to evaluate the physics-informed loss terms via  automatic differentiation in TensorFlow \citep{tensorflow2015-whitepaper}. Then, starting with guesses $K_c^{(0)}$ and $K_{\eta}^{(0)}$ for the model parameters, we calculate the loss terms corresponding to the particle transport and momentum equations, as well as any constraints. The activation function is the hyperbolic tangent function. During the process of minimizing $\mathscr{L}$, $\bm{\Theta}_u^{(k)}$, $\bm{\Theta}_\phi^{(k)}$, $K_c^{(k)}$ and $K_\eta^{(k)}$ are updated at each iteration $k$. The loss function is minimized first using ``Adam'' \citep{KB15}, which is a stochastic gradient descent algorithm, and  then with ``L-BFGS-B'' subsequently. The stopping criterion for the optimization is that the change in the loss function between iterations is less than machine precision. However, this stopping criterion may or may not satisfy our convergence criterion. So, upon the stoppage of the optimization procedure, we check that  $\mathscr{L}<\mathrm{TOL}$, for some prescribed tolerance $\mathrm{TOL}\simeq 10^{-2}$. Upon satisfaction of the latter criterion, we consider the solution converged. Then, we have obtained not only the optimized neural networks' weights $\bm{\Theta}_u$ and $\bm{\Theta}_\phi$, but also  the initially unknown model parameters $K_c$ and $K_\eta$.

\section{Couette flow}
\label{sec:couette}
\subsection{Governing physics equations}
\label{eq:eqs_Couette}

For flow between concentric rotating cylinders, the domain is $\mathcal{V}=\{\bm{x}=(r,\theta) \;|\; \kappa R\le r\le R,\; 0\le \theta<2\pi\}$, where $\kappa<1$ is dimensionless, and the problem is independent of the axial, $z$, coordinate. The unidirectional velocity field is thus $\bm{u} = u_\theta(r)\bm{e}_\theta$. The radial component of the momentum equation~\eqref{eq:mom_steady} (now in cylindrical coordinates \citep{Panton}) reduces to
\begin{equation}
    \frac{1}{r^2}\frac{\partial}{\partial r} \left( r^2\tau_{r\theta}\right) = 0 \qquad\Rightarrow\qquad \tau_{r\theta} = \frac{A}{r^2},
\end{equation}
where $A$ is a constant to be calculated by imposing boundary conditions \cite{pabga92}. Note that in this case of axisymmetric flow, $\partial p/\partial \theta=0$. 
Equivalently,   
\begin{equation}
    \dot{\gamma} = \dot{\gamma}_{r\theta} = \frac{A}{r^2\eta},\qquad A = \frac{-\Omega}{\int_{\kappa R}^R (r^3\eta)^{-1} \,dr},
    \label{eq:mom_Couette}
\end{equation}    
where $\Omega$ is the angular velocity of the inner cylinder. On dimensional grounds alone, for a neutrally buoyant non-Brownian suspension at low Reynolds number at steady state, it is expected that $\eta/\eta_s = \eta_r(\phi)$ \citep{SP05}, where $\eta_s$ is the Newtonian solvent's dynamic viscosity, and $\eta_r(\phi)$ is the dimensionless contribution to the viscosity due to particles. Indeed, the experiments of \citet{pabga92} were verified to be in this specific regime, and they used the empirical Krieger--Dougherty relation:
\begin{equation}
    \eta_r(\phi) = \left(1 - \frac{\phi}{\phi_m} \right)^{-a},
    \label{eq:vis_Krieger}
\end{equation}
where $\phi_m = 0.68$ is the maximum packing volume fraction, and $a=1.82$ is a positive empirical exponent (in principle, related to $\phi_m$ \citep{SP05}). The form of Eq.~\eqref{eq:vis_Krieger} and values for $\phi_m$ and $a$ are based on experimental correlations \citep{KD59, Krieger72}, and they are well-established for non-Brownian suspensions \citep{DM14,GuaMor}. Here, we are only interested in interrogating the particle migration model, so any fitting parameters for the fluid and suspension properties are taken as per the literature.

Next, dimensionless governing equations are obtained by scaling the dimensional variables in the following way: \begin{equation}
    r^* = r/R,\qquad 
    \dot{\gamma}^*(r^*) = \dot{\gamma}_{r\theta}(r)/\Omega,\qquad 
    u^*(r^*) = u_{\theta}(r)/U_\mathrm{max},
    \label{eq:couette_nondimensional}
\end{equation}
where $U_\mathrm{max} = \Omega \kappa R$ is the maximum velocity at the rotating wall.
Therefore, the momentum equation~\eqref{eq:mom_Couette} can be written as
\begin{equation}
    \dot{\gamma}^*(r^*) = \frac{-1}{r^{*2}\eta_r\int_{\kappa}^1 (r^{*3}\eta_r)^{-1} \,dr^*}.
    \label{eq:couette_momentum}
\end{equation}

Neglecting the Brownian flux, the corresponding dimensionless particle transport equation~\eqref{eq:steady_susp} for the non-Brownian suspension in this geometry is
\begin{equation}
    \frac{1}{\dot{\gamma}^*}\frac{d\dot{\gamma}^*}{dr^*} + \frac{1}{\phi}\frac{d\phi}{dr^*} + \frac{K_{\eta}}{K_c}\frac{1}{\eta_r}\frac{d\eta_r}{dr^*} = 0.
    \label{eq:couette_susp}
\end{equation}
Observe that in this (non-Brownian) case, only the ratio $K_\eta/K_c$ of the two model parameters appears up in the final form of the particle transport equation. In other words, there is only a single quantity to ``fit,'' and this fact is reflected in the PINN architecture used for this problem.

Additionally, the particle distribution satisfies 
\begin{equation}\label{eq:phib_int}
    \frac{1}{1-\kappa}\int_{\kappa}^1 \phi(r^*) \,dr^* = \phi_b,
\end{equation}
where $\phi_b$ is the bulk volume fraction. The uniform distribution of particles at the initial time is then $\phi(r,t=0)=\phi_b$, and at steady state conservation of mass requires that Eq.~\eqref{eq:phib_int} hold.

\citet{pabga92} suggested, based on analysis of their experiments, that $K_c/K_{\eta} \approx 0.66$ best fits the profile measured for steady-state Couette flow of suspensions of $2a_p=675$ \si{\micro\meter} particles at $\phi_b = 0.55$. The agreement was also good when comparing with experimental data at $\phi_b =0.45$, $0.50$, and $0.55$. In this section, we wish to investigate the best-fit value of $K_c/K_{\eta}$ obtained by the PINN approach for solving the inverse problem.

\subsection{PINN loss function}
For this problem, the loss function is 
\begin{equation}
    \mathscr{L} = w_u \MSE_u + w_{\phi} \MSE_{\phi} + w_p \MSE_p + w_m \MSE_m 
    + w_c \MSE_c.
    \label{eq:loss_Couette}
\end{equation}
The notation for the $\MSE$ terms was introduced in Sec.~\ref{sec:PINN_algo}. These terms are now implemented as  
\begin{subequations}\begin{align}
\MSE_u &= \frac{1}{N}\sum_{i = 1}^{N} |u^* - u_\mathrm{train}^*|^2 ,\\ 
\MSE_{\phi}  &= \frac{1}{N}\sum_{i = 1}^{N} |\phi - \phi_\mathrm{train}|^2, \\
\MSE_m &= \frac{1}{N}\sum_{i = 1}^{N}\left|\dot{\gamma}^* + \frac{1}{r^{*2}\eta_r\int_{\kappa}^1 (r^{*3}\eta_r)^{-1} \,dr^*}\right|^2, \label{eq:Couette_MSE_m}\\
\MSE_p &= \frac{1}{N}\sum_{i = 1}^{N}\left|\frac{1}{\dot{\gamma}^*}\frac{d\dot{\gamma}^*}{dr^*} + \frac{1}{\phi}\frac{d\phi}{dr^*} + \frac{K_{\eta}}{K_c}\frac{1}{\eta_r}\frac{d\eta_r}{dr^*}\right|^2,\label{eq:Couette_MSE_s}\\
\MSE_c &=  \left|\frac{1}{1-\kappa}\int_{\kappa}^1 \phi(r^*)\, dr^* - \phi_b\right|^2. \label{eq:Couette_MSE_phib}
\end{align}
\end{subequations}
The integrals in the expressions in Eqs.~\eqref{eq:Couette_MSE_m} and \eqref{eq:Couette_MSE_phib} are approximated by averages over the randomly selected set of collocation points (essentially a Monte Carlo quadrature): $\int_\kappa^1 (\cdot) = \sum_{i = 1}^{N} (\cdot)\times (1-\kappa)/N$. Appendix~\ref{sec:pre-proc} describes how the experimental data (digitized from \citep{pabga92}) was pre-processed into training data.

Note that the output variables of the NNs are $u$ and $\phi$, as depicted in Fig.~\ref{fig:pinn_diagram}. Thus, the remaining variables, $\eta_r$ and $\dot{\gamma}^*$, have to be expressed in terms of these output variables. Recall that $\dot{\gamma} = r d(u_{\theta}/r)/dr$ for purely azimuthal flow by definition \cite{Panton}. Then, using the nondimensionalization from Eq.~\eqref{eq:couette_nondimensional}:
\begin{equation}
    \dot{\gamma}^*(r^*) 
    =\kappa r^*\frac{d}{dr^*}\bigg(\frac{u^*(r^*)}{r^*}\bigg).
    \label{eq:dot_gamma}
\end{equation}
Substituting the expression for $\dot{\gamma}^*$ from Eq.~\eqref{eq:dot_gamma} and $\eta_r$ from Eq.~\eqref{eq:vis_Krieger} into the MSE expressions \eqref{eq:Couette_MSE_m} and \eqref{eq:Couette_MSE_s}, the loss function from Eq.~\eqref{eq:loss_Couette} now depends only on the neural networks' weights ($\bm{\Theta}_u$ and $\bm{\Theta}_\phi$) and the ratio $K_c/K_\eta$. By minimizing the resulting $\mathscr{L}(\bm{\Theta}_u, \bm{\Theta}_\phi, K_c/K_\eta)$ with respect to its arguments, we obtain $\NN(u)$, $\NN(\phi)$, and the value of $K_c/K_\eta$ that simultaneously lead to the best agreement with the training data \emph{and} the physics.

\subsection{Comparison between PINN, theory, and experiment}
\label{sec:couette_comparison}

To evaluate the ``theoretical" particle distribution profile, we use $d\eta_r/dr^* = (d\eta_r/d\phi)(d\phi/dr^*)$ to re-write Eq.~\eqref{eq:couette_susp} as:
\begin{equation}
    \left( \frac{1}{\phi} + \frac{K_{\eta}}{K_c}\frac{1}{\eta_r}\frac{d\eta_r}{d\phi}\right)\frac{d\phi}{dr^*} = - \frac{1}{\dot{\gamma}^*}\frac{d\dot{\gamma}^*}{dr^*} = \left(\frac{2}{r^*} + \frac{1}{\eta_r}\frac{d\eta_r}{d\phi}\frac{d\phi}{d r^*}\right),
    \label{eq:couette_susp_ode_1}
\end{equation}
where the second equality follows from using the dimensionless version of Eq.~\eqref{eq:mom_Couette}. Now, Eq.~\eqref{eq:couette_susp_ode_1} can be rearranged as an ODE for $\phi(r^*)$:
\begin{equation}
    \left[ \frac{1}{\phi} + \left(\frac{K_{\eta}}{K_c}-1\right)\frac{1}{\eta_r}\frac{d\eta_r}{d\phi} \right]\frac{d\phi}{dr^*} = \frac{2}{r^*}.
    \label{eq:couette_susp_ode_2}
\end{equation}
Finally, we can use the Krieger viscosity from Eq.~\eqref{eq:vis_Krieger}, multiply both sides of Eq.~\eqref{eq:couette_susp_ode_2} by $\phi$ and solve for $d\phi/dr^*$, to obtain a first-order nonlinear ODE for $\phi(r^*)$:
\begin{equation}
    \frac{d\phi}{dr^*} = \frac{2\phi }{[(K_\eta/K_c-1)a(\phi/\phi_m)(1 - \phi/\phi_m)^{-1} + 1 ]r^*},
    \label{eq:couette_susp_ode_3}
\end{equation}
which is the same as Eq.~(21) of \citet{pabga92}. Equation~\eqref{eq:couette_susp_ode_3} can be integrated from $r^*=\kappa$ to $r^*=1$ using an arbitrary value $\phi(r^*=\kappa) = \phi_w \in [0,1]$ as the initial condition. Then, a nonlinear iteration (implemented using \texttt{optimize.root\_scalar} from the SciPy stack in Python \citep{SciPy}) updates $\phi_w$ until the constrain~\eqref{eq:phib_int} is satisfied for the given $\phi_b$. This solution will be shown as the ``theory'' curve in the figures below.

In Fig.~\ref{fig:couette}, we compare our PINN results using $w_u = w_\phi = w_m = w_p = w_c = 1$ to theory (numerical solution of Eq.~\eqref{eq:couette_susp_ode_3}) and the experiments (by \citet{pabga92}). We use neural networks with two hidden layers with 10 nodes in each layer and a learning rate of 0.001 for the Adam optimizer. The process of choosing a suitable number of layers and nodes is discussed in  Appendix~\ref{sec:choice_layers}.

Two versions of the theoretical prediction for the particle distribution profile are shown in Fig.~\ref{fig:couette}(b). One is calculated by solving Eq.~\eqref{eq:couette_susp_ode_3} (numerically by the method described above) using the value $K_c/K_\eta = 0.66$ motivated in \citep{pabga92}. The other theoretical prediction is derived via the \textit{ad-hoc} approximation $1.82(1-K_\eta/K_c)\approx-1$ made by  \citet{pabga92}. We observe that the PINN prediction is an improvement over the numerical solution of Eq.~\eqref{eq:couette_susp_ode_3}. Surprisingly, the approximation $1.82(1-K_\eta/K_c)\approx-1$ improves the agreement further between the theory and experiments. This approximation, made out of convenience in \citep{pabga92}, does not appear to be justifiable on the basis of mathematical grounds (such as perturbation theory), therefore its good agreement with the experiment must be simply coincidence.

\begin{figure}
  \centering
  \includegraphics[width=0.85\textwidth]{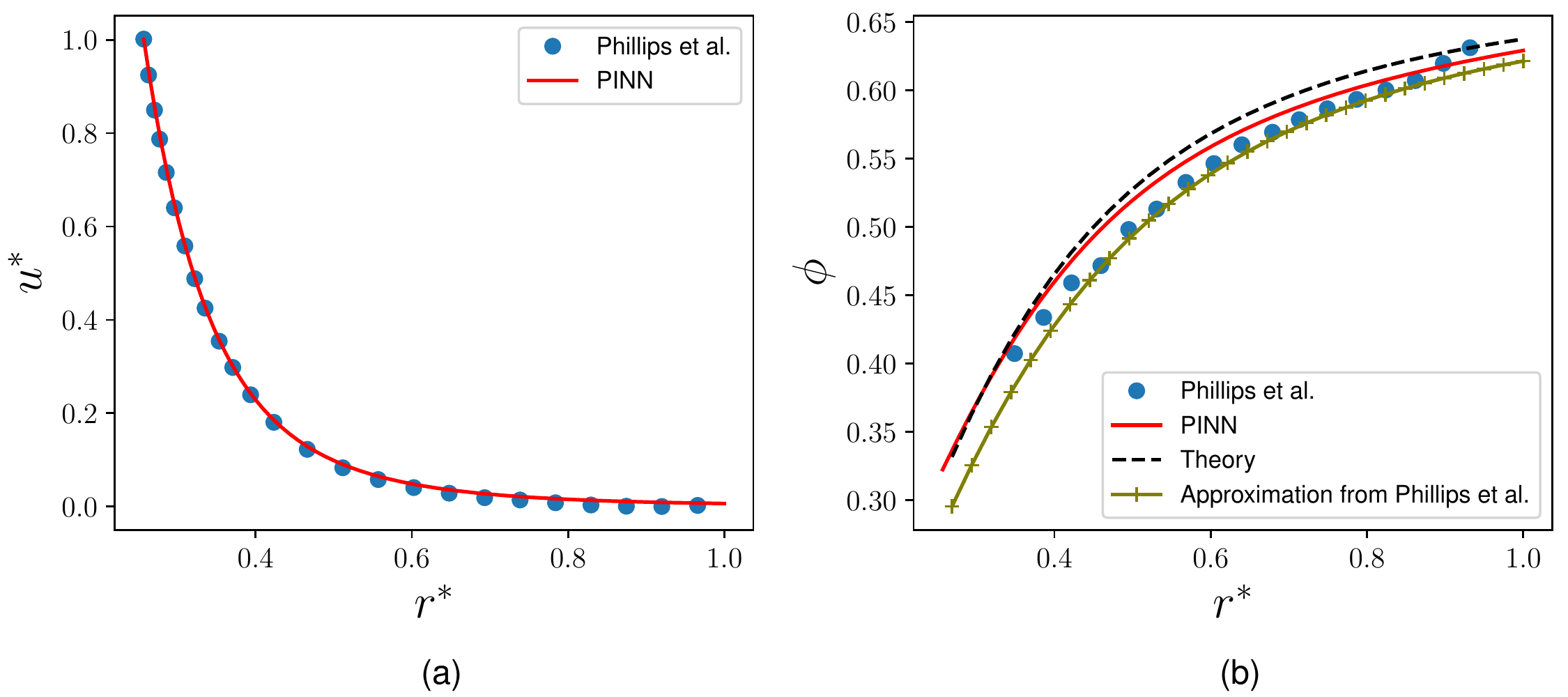}
\caption{Validation of the proposed PINN approach to shear-induced migration. The PINN is applied to analyze the experiment from \citet{pabga92} in a concentric Couette cell at $\phi_b = 0.55$, yielding: (a) the dimensionless velocity profile $u^*(r^*)$ and (b) the particle distribution (volume fraction) $\phi(r^*)$. Symbols represent the experimental data from \citet{pabga92}; red solid curves are PINN predictions; the dashed curve is the numerical solution of Eq.~\eqref{eq:couette_susp_ode_3}; the curve with cross symbols in (b) is the approximate analytical solution from \citet[Eqs.~(24)--(25)]{pabga92}. The PINN ``learns'' a value of the unknown model parameter $K_c/K_\eta \approx 0.66$, which is in agreement with \citep{pabga92}.}
\label{fig:couette}
\end{figure}

As evidenced by Fig.~\ref{fig:couette}, we obtain good agreement between the PINN predictions and the experimental data, for both  $u^*$ and $\phi$. To account for the statistical variation in the converged (``learned'') value of $K_c/K_\eta$ due to the random initialization of the NNs, we averaged the predictions from 1000 different initializations (excluding cases that failed to train, $\approx 11\%$ of the total, in which the loss was not minimized to the given tolerance) to obtain a statistical result with mean and standard error: $K_c/K_\eta = 0.66 \pm 0.05 $, which is consistent with the fitted value in \citep{pabga92} by their non-machine-learning approach. Thus, we have validated the PINN approach for the shear-induced migration of non-Brownian particles in a concentric Couette cell, showing that PINNs not only provide suitable predictions for the velocity and particle distribution profiles but also ``learn'' the accepted value of the model parameter $K_c/K_\eta$ given in the literature.

\section{Poiseuille flow}
\label{sec:Poiseuille}

\subsection{Governing physics equations}

For Poiseuille flow in a slot of height $2H$, the domain is $\mathcal{V}=\{\bm{x}=(x,y)\;|\; -\infty<x<+\infty,\, -H\le y\le +H\}$. For fully developed flow, the $x$ dependence drops out. The unidirectional velocity field is thus $\bm{u} = u_x(y)\bm{e}_x$. We introduce the dimensionless variables
\begin{equation}
    x^* = x/L,\qquad 
    y^* = y/H,\qquad
    \dot{\gamma}^*(y^*) = \dot{\gamma}(y)/\dot{\gamma}_0,\qquad
    u^*(y^*) = u_x(y)/U_\mathrm{max},\qquad
    \label{eq:pois_nondimensional}
\end{equation}
where $U_\mathrm{max}$ is the centerline (maximum) velocity of the solvent fluid at the same volumetric flow rate, $\dot{\gamma}_0 = U_\mathrm{max}/H $ is the mean shear rate, and $L$ is a typical axial length scale for the channel.

Now, the dimensionless particle transport equation~\eqref{eq:steady_susp} for the suspension takes the form
\begin{equation}
    K_c \phi \left(\phi \frac{d^2u^*}{dy^{*2}} + \frac{du^*}{dy^*} \frac{d\phi}{dy^*} \right) + K_{\eta} \frac{du^*}{dy^*}\frac{\phi^2}{\eta_r}\frac{d\eta_r}{dy^*} + \frac{1}{Pe}\frac{d\phi}{dy^*}=0.
    \label{eq:pois_susp2}
\end{equation}
In Eq.~\eqref{eq:pois_susp2}, $Pe$ is the shear P\'eclet number, which quantifies the relative importance of shear migration to Brownian migration of particles \citep{SP05}, defined as 
\begin{equation}
    Pe = \frac{a_p^2\dot{\gamma}_0}{D} = \frac{6\pi \eta_sa_p^3\dot{\gamma_0}}{k_B T},
    \label{eq:Pe_defn}
\end{equation}
where $k_B$ is Boltzmann's constant, and $T$ is temperature.
Observe that unlike the case of Eq.~\eqref{eq:couette_susp}, Eq.~\eqref{eq:pois_susp2} for finite $Pe$ cannot be divided by $K_c$ (to only consider the ratio $K_\eta/K_c$). Again, the particle distribution is constrained such that 
\begin{equation}
    \frac{1}{2}\int_{-1}^{+1} \phi(y^*) \,dy^* = \int_0^1 \phi(y^*) \,dy^* = \phi_b.
    \label{eq:phi_b_integral}
\end{equation}

For a dense non-Brownian suspension ($Pe\gg1$), the velocity is not parabolic \citep{pabga92,KHL1994}. Its shape is found by solving the momentum equation~\eqref{eq:mom_steady} for the pressure-driven Poiseuille flow of the suspension:
\begin{equation}
    \frac{d}{dy^*}\bigg(\eta_r(\phi) \frac{du^*}{dy^*} \bigg) = G^* = \frac{GH^2}{\eta_s U_\mathrm{max}} ,
    \label{eq:momentum_poiseuille}
\end{equation}
where $G^*$ (resp.~$G$) is the dimensionless (resp.~dimensional) axial pressure gradient, which is constant in unidirectional flow \cite{Panton}. Integrating Eq.~\eqref{eq:momentum_poiseuille} once and imposing a centerline symmetry condition, we have
\begin{equation}
    \eta_r(\phi) \frac{du^*}{dy^*} = G^* y^*.
    \label{eq:momentum_poiseuille2}
\end{equation}
Similarly to the approach of \citet{RHPT20}, Eq.~\eqref{eq:momentum_poiseuille2} will be enforced via the PINN's loss function to account for the blunted (non-parabolic) velocity profiles of dense suspensions. However, we will not enforce no-slip boundary conditions with Eq.~\eqref{eq:momentum_poiseuille2} because experiments \citep{KHL1994,JKA95,K05} suggest that dense suspensions can slip along the channel walls (see Fig.~\ref{fig:koh_11}). The proposed machine learning methodology naturally handles slip without further effort. To calculate $G^*$ from the experimental data, Eq.~\eqref{eq:momentum_poiseuille2} is integrated from $y^*=0$ to $y^*=1$ and rearranged, yielding 
\begin{equation}
G^* = \frac{u^*(1)-u^*(0)}{\int_{0}^{1}y^*/\eta_r(\phi)\,dy^*},
\label{eq:Gstar}
\end{equation}
which is evaluated using the experimental $u^*$ and $\phi$ profiles and the trapezoidal rule for the integral.

For a strongly Brownian suspension ($Pe=\mathcal{O}(1)$), as we will discuss in Sec.~\ref{sec:Brownian}, the velocity profile in experiments \citep{FAWM03} is indistinguishable from a parabolic one, so instead of Eq.~\eqref{eq:momentum_poiseuille2}, we can simply enforce
\begin{equation}
    u^*(y^*) = 1 - y^{*2}.
    \label{eq:momentum_poiseuille_parabolic}
\end{equation}
Put differently: now $\eta_r(\phi)\approx 1$, $u^*(1)=0$ (the dilute suspension does not slip), and the scale $U_\mathrm{max}$ is chosen to make $G^*=-2$ in this case, consistent with Eq.~\eqref{eq:Gstar}.

\subsection{Non-Brownian dense suspension}
\label{sec:Poiseuille_NB}

\subsubsection{PINN loss function}
In a non-Brownian dense suspension, the Brownian diffusive flux $\bm{N}_B$ can be neglected in Eq.~\eqref{eq:steady_susp}, which eliminates the term $Pe^{-1}d\phi/dy^*$ from Eq.~\eqref{eq:pois_susp2} (equivalently, the limit $Pe\to\infty$ corresponds to a non-Brownian suspension). 
Then, in this case, the $\MSE$ terms as introduced in Eq.~\eqref{eq:loss} in Sec.~\ref{sec:PINN_algo} are now be implemented as:
\begin{subequations}\begin{align}
\MSE_u &= \frac{1}{N}\sum_{i = 1}^{N} |u^* - u_\mathrm{train}^*|^2 , \\
\MSE_{\phi}  &= \frac{1}{N}\sum_{i = 1}^{N} |\phi - \phi_\mathrm{train}|^2, \\
\MSE_m &= \frac{1}{N}\sum_{i = 1}^{N}\left|\eta_r(\phi) \frac{du^*}{dy^{*}} - G^*y^*\right|^2,\\
\MSE_p &= \frac{1}{N}\sum_{i = 1}^{N}\left|\frac{K_c}{K_{\eta}} \phi \left(\phi \frac{d^2u^*}{dy^{*2}} + \frac{du^*}{dy^*} \frac{d\phi}{dy^*} \right) +  \frac{du^*}{dy^*}\frac{\phi^2}{\eta_r}\frac{d\eta_r}{dy^*}\right|^2, \\
\MSE_c  &=\left|\int_0^1 \phi(y^*) dy^* - \phi_b\right|^2.
\end{align}\label{eq:MSE_nonBrownian}\end{subequations}
Appendix~\ref{sec:pre-proc} describes how the experimental data (digitized from \citep{KHL1994}) was pre-processed into training data.
 
Note that in $\MSE_p$, $ {d\eta_r}/{dy^*}$ is calculated by the chain rule: ${d\eta_r}/{dy^*} =  ({d\eta_r}/{d\phi})({d\phi}/{dy^*})$. Then, substituting $\eta_r$ from Eq.~\eqref{eq:vis_Krieger} into the loss function, we obtain $\mathscr{L}$,
which depends only on the neural networks' weights ($\bm{\Theta}_u$ and $\bm{\Theta}_\phi$) and the ratio $K_c/K_\eta$. By minimizing the resulting $\mathscr{L}(\bm{\Theta}_u, \bm{\Theta}_\phi, K_c/K_\eta)$ with respect to its arguments, and using $w_u = w_\phi= w_m = w_p = w_c = 1$, we find suitable neural networks' weights and the value of the parameter $K_c/K_\eta$ that simultaneously leads to the best agreement with the training data \emph{and} the physics.

\subsubsection{Comparison between PINN, theory and experiment}
\label{sec:comparison_koh}

We can solve for the ``theoretical'' prediction for $\phi$ from Eq.~\eqref{eq:pois_susp2}. For a non-Brownian suspension, the dependence of the viscosity on the particle volume fraction is given by Eq.~\eqref{eq:vis_Krieger}, and the velocity profile obeys Eq.~\eqref{eq:momentum_poiseuille2}. Substituting these expressions into Eq.~\eqref{eq:pois_susp2} and neglecting the Brownian term ($Pe\to\infty$), we once again obtain a nonlinear first-order ODE for $\phi(y^*)$:
 \begin{equation}
     \frac{d\phi}{dy^*} =\frac{\phi}{\left[ \left(1-{K_{\eta}}/{K_c} \right) a(\phi/\phi_m) \left(1 - {\phi}/{\phi_m} \right)^{-1} - 1\right] y^*},
     \label{eq:ode_koh_1}
\end{equation}
Then, Eq.~\eqref{eq:ode_koh_1} can be integrated numerically from $y^*=1$ back to $y^*=0$ (to handle the singularity at $y^*=0$). An arbitrary value $\phi(y^*=1) = \phi_w\in [0,1]$ is used as an initial guess. Then, a nonlinear iteration (implemented using \texttt{optimize.root\_scalar} from the SciPy stack in Python \citep{SciPy}) updates $\phi_w$ until the constrain~\eqref{eq:phi_b_integral} is satisfied for the given $\phi_b$. This solution will be shown as the ``theory'' curve in the figures below. Note that, while 
\citet{KHL1994} assumed $K_c/K_{\eta} = 0.66$ (based on the result in Sec.~\ref{sec:couette_comparison}), we use the value of $K_c/K_{\eta}$ learned by the PINN instead. 

\begin{figure}[ht!]
  \centering
  \includegraphics[width=0.7\textwidth]{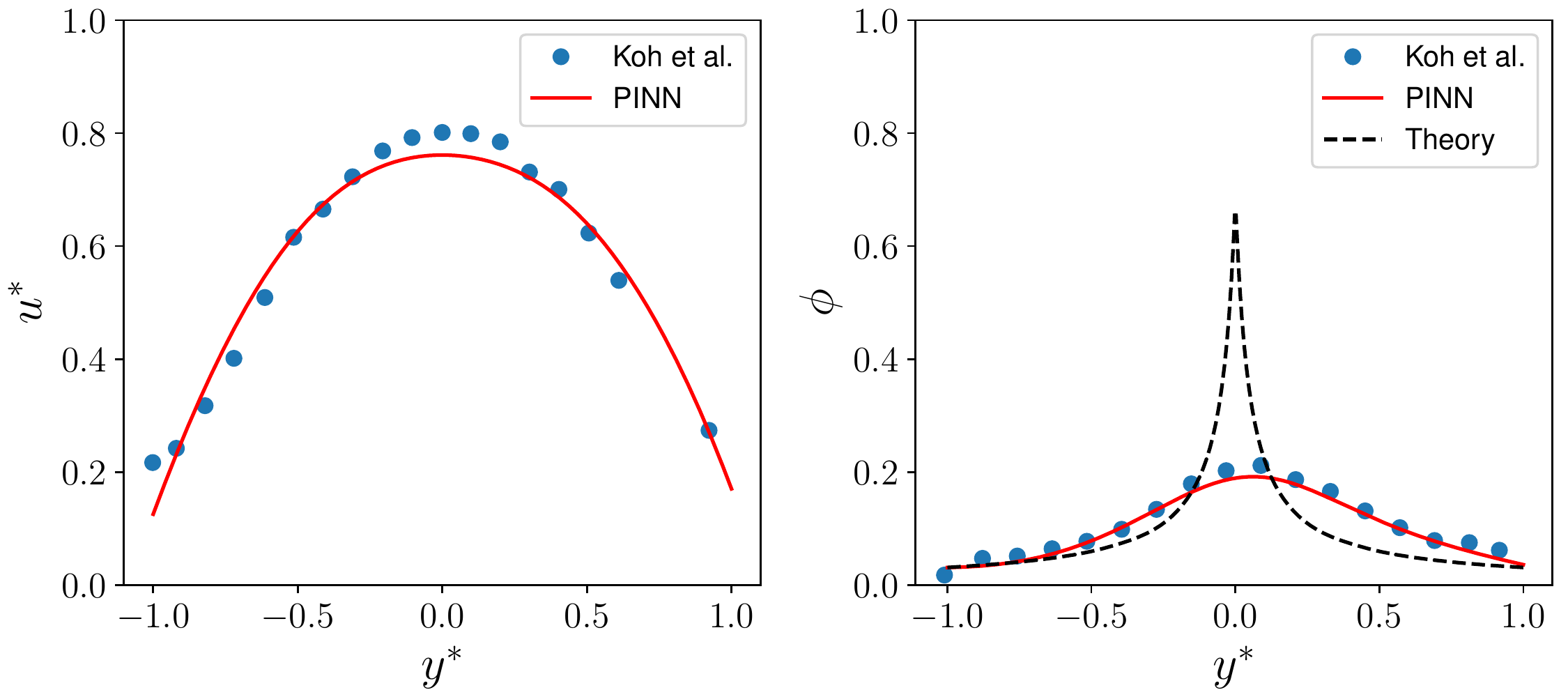}
\caption{Application of the proposed PINN-based approach for understanding shear-induced migration to experiment 187 from \citet{KHL1994} at $\phi_b = 0.1$ and $G^*=-1.51$. The PINN found $K_c/K_{\eta} = 0.10 \pm 0.012$.}
\label{fig:koh_11}
\end{figure}

\begin{figure}[ht!]
  \centering
  \includegraphics[width=0.7\textwidth]{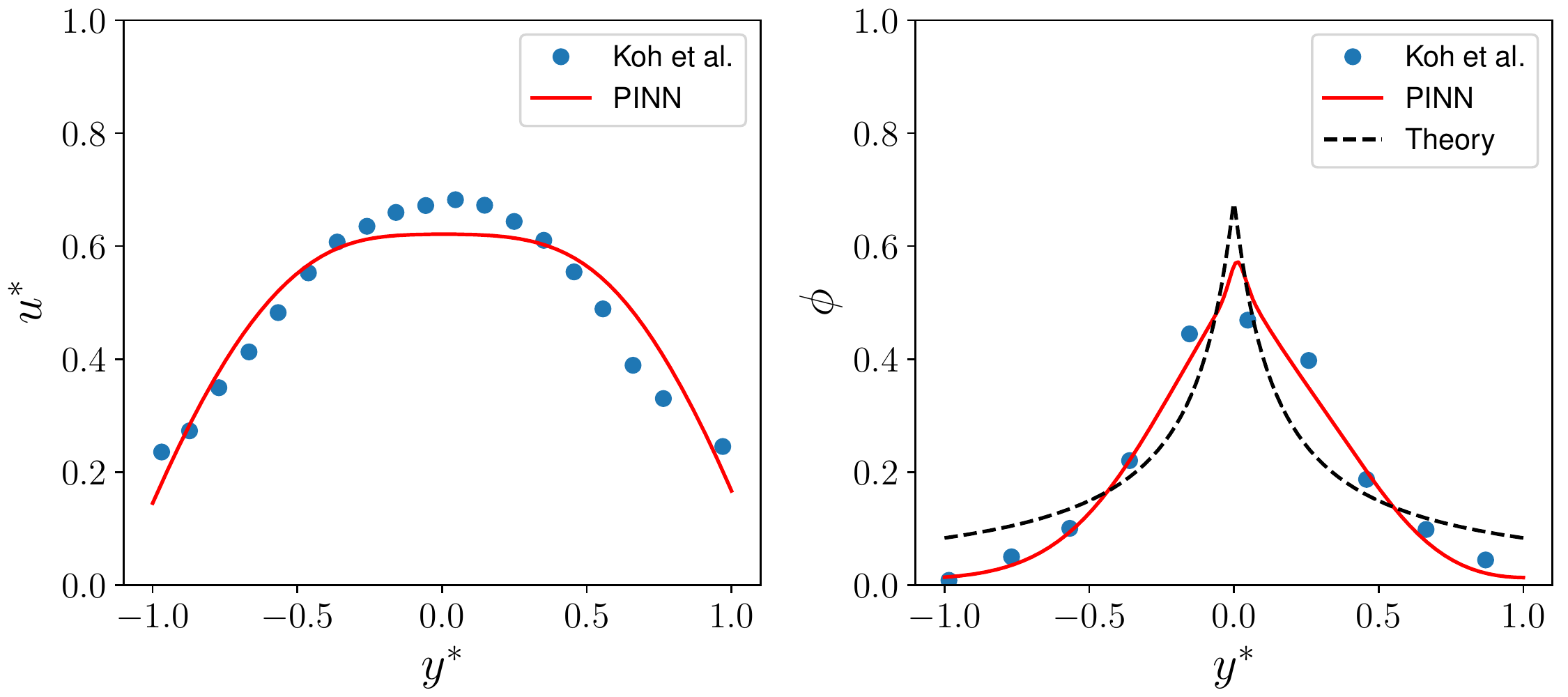}
\caption{Application of the proposed PINN-based approach for understanding shear-induced migration to experiment 189 from \citet{KHL1994} at $\phi_b = 0.2$ and $G^*=-1.22$. The PINN found $K_c/K_{\eta} = 0.44 \pm 0.018$.}
\label{fig:koh_15}
\end{figure}

\begin{figure}[ht!]
  \centering
  \includegraphics[width=0.7\textwidth]{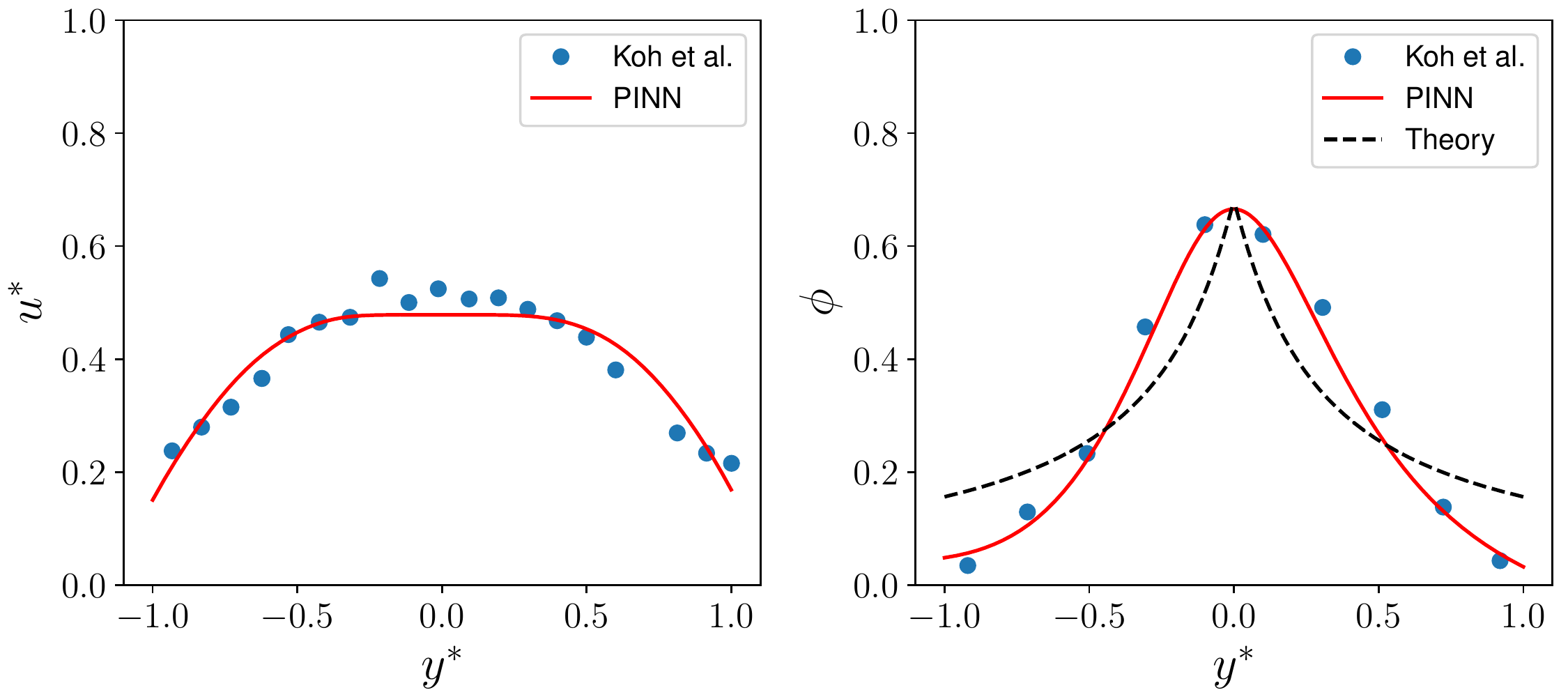}
\caption{Application of the proposed PINN-based approach for understanding shear-induced migration to experiment 192 from \citet{KHL1994} at $\phi_b = 0.3$ and $G^*=-1.04$. The PINN found $K_c/K_{\eta} = 0.58 \pm 0.020$.}
\label{fig:koh_19}
\end{figure}

In Fig.~\ref{fig:koh_11}, \ref{fig:koh_15}, \ref{fig:koh_19} we compare the PINN solutions to the laser-Doppler velocimetry experimental measurements of \citet{KHL1994} at $\phi_b = 0.1, 0.2, 0.3$, respectively.  As in Sec.~\ref{eq:eqs_Couette}, we use the scaled viscosity function $\eta_r(\phi)$ given in Eq.~\eqref{eq:vis_Krieger}, because their experiments are also for a neutrally buoyant non-Brownian suspension at low Reynolds number,
like \citep{pabga92}. As can be deduced from the figures, the PINN predictions for the profiles $u^*(y^*)$ and $\phi(y^*)$ agree well with the experiment data. Specifically, the PINN predicts $\phi$ better than pure theory from Eq.~\eqref{eq:ode_koh_1} because, when using Eq.~\eqref{eq:ode_koh_1} as a constraint on the learning process, the PINN ``smooths out'' the physically-questionable singularity of the ODE at $y^*=0$. Note that this feature of the PINN approach was also mentioned in \cite{RHPT20}, in the context of the shear stress singularity at the channel centerline under a power-law rheological model.

\begin{table}[ht]
\centering
\begin{tabular}{l|l}
\hline
\hline
$\phi_b$ & $K_c/K_\eta$\\
\hline
$0.1$ & $0.10 \pm 0.012$\\
$0.2$ & $0.44 \pm 0.018$\\
$0.3$ & $0.58 \pm 0.020$\\ \hline
\hline
\end{tabular}
\caption{Values of the shear-induced migration model's parameter $K_c/K_\eta$,  as inferred by the PINN from non-Brownian experimental data of \citet{KHL1994}, for different bulk volume fractions $\phi_b$. As before, these statistical results, with a mean and a standard error, come from 1000 runs of the PINN algorithm using different random initializations of the NNs (excluding cases that failed to train).}
\label{tab:koh_params}
\end{table}


Importantly, by training the PINN, we deduce best-fit $K_c/K_\eta$ values \emph{different} from the traditional value of $0.66$, which has only been validated for the concentric Couette flow (Sec.~\ref{sec:couette_comparison}). Table~\ref{tab:koh_params} summarizes the values that the PINN ``learns'' from experimental data with different values of $\phi_b$ for the Poiseuille flow in a slot.

\subsection{Brownian suspension}
\label{sec:Brownian}

Now, we return to the ``full'' Eq.~\eqref{eq:pois_susp2} at finite $Pe$, which was defined in Eq.~\eqref{eq:Pe_defn}. Now, it is expected that $\eta/\eta_s=\eta_r(\phi,Pe)$ \citep{SP05} (see also \citep[Ch.~7]{GuaMor}), while the suspension is still neutrally buoyant, at steady state and at low Reynolds number. (The geometry is still that of Poiseuille flow in a slot as introduced in Sec.~\ref{sec:Poiseuille_NB}.) 

Unlike the previous sections, we can no longer use the Krieger--Dougherty viscosity function from Eq.~\eqref{eq:vis_Krieger} because the Brownian suspension is expected to be shear-thinning over a wide range of shear rates when the dependence on $Pe$ is accounted for \cite{SP05}. Motivated by the work of \citet{KM20}, we take the Brownian suspension's shear viscosity to be
\begin{equation}
    \eta_r(\phi,Pe) = \eta_{\infty}(\phi) + \frac{\eta_0(\phi)-\eta_{\infty}(\phi)}{1+K_{Pe} \eta_sa_p^3\dot{\gamma}/(k_B T)}
    = \eta_{\infty}(\phi) + \frac{\eta_0(\phi)-\eta_{\infty}(\phi)}{1+K_{Pe} Pe\dot{\gamma}^*/(6\pi)},
    \label{eq:vis_kang}
\end{equation}
where
\begin{subequations}
\begin{align}
    \eta_0(\phi) &= (1-\phi/\phi_{m_0})^{-\mathfrak{a}_0},\\
    \eta_{\infty}(\phi) &= (1-\phi/\phi_{m_\infty})^{-\mathfrak{a}_\infty},
\end{align}
\end{subequations}
based on the correlations proposed by \citet{KIV85}. Typically, $\phi_{m_0} = 0.63$, $\phi_{m_\infty} = 0.71$, $\mathfrak{a}_0=1.96$, $\mathfrak{a}_\infty=1.93$, and $K_{Pe} = 1.31$ are used in the literature \citep{KM20}  based on experimental fits. While the zero-$Pe$ and infinite-$Pe$ ``plateaus'' of the viscosity function~\eqref{eq:vis_kang} can be measured accurately (yielding the maximum volume fractions $\phi_{m_0}$ and $\phi_{m_\infty}$, along with the exponents $\mathfrak{a}_0$ and $\mathfrak{a}_\infty$), the transition over intermediate $Pe$ is characterized by the dimensionless parameter $K_{Pe}$. This parameter is harder to infer from a single experiment (and, indeed, has \emph{not} been reported as being independently measured in the experimental papers on shear-induced migration of Brownian particles), thus we propose to treat it as \textit{a priori} unknown, like $K_c$ and $K_\eta$. In other words, we will self-consistently determine the unknown $K_{Pe}$ via the inverse formulation in the PINN approach applied to the experiments of \citet{FAWM03} on shear-induced migration of colloidal particles.

\begin{figure}
  \centering
  \includegraphics[width=0.5\textwidth]{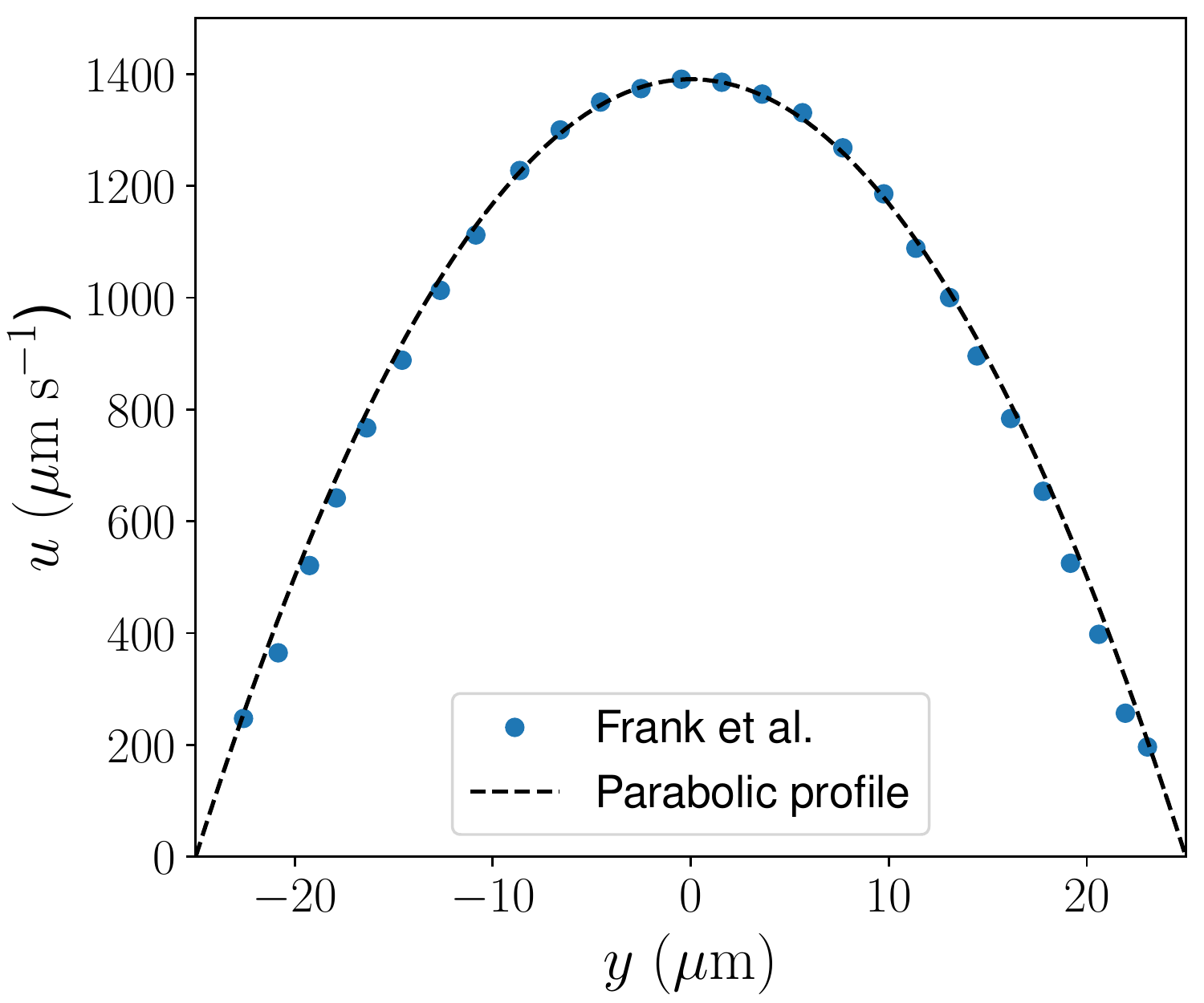}
\caption{Measured axial velocity of a Brownian suspension in Poiseuille flow reproduced from \citet{FAWM03} is compared to a dimensional version of the parabolic profile from Eq.~\eqref{eq:momentum_poiseuille_parabolic}, showing good agreement. The channel width is $2H=50$ \si{\micro\meter}, and $U_\mathrm{max} \approx 1391$ \si{\micro\meter \per \second} from the inset of Fig.~3 of in \citep{FAWM03}.}
\label{fig:frank_vel}
\end{figure}

Observe that we keep the variable local shear rate $\dot{\gamma}^*=\dot{\gamma}^*(y^*) = du^*/dy^*$ (dimensionless, recall Eq.~\eqref{eq:pois_nondimensional}) in Eq.~\eqref{eq:vis_kang} because $\dot{\gamma}^*\ne const.$ in Poiseuille flow. Nevertheless, using particle-image velocimetry, \citet{FAWM03} found experimentally that the velocity profile of their Brownian suspension (of $2a_p=2$ \si{\micro\meter} spherical colloidal particles) flowing through a rectangular channel only slightly deviates from the parabolic profile of the solvent (see Fig.~\ref{fig:frank_vel}), hence $\dot{\gamma}^*(y^*)\approx -2y^*$. Therefore, for the Brownian suspensions, we shall use the parabolic profile from  Eq.~\eqref{eq:momentum_poiseuille_parabolic} to define $\MSE_m$ in the loss function, instead of the full momentum equation.

\subsubsection{PINN loss function}
For the Brownian suspension, the $\MSE$ terms are as introduced in Eq.~\eqref{eq:loss} in Sec.~\ref{sec:PINN_algo} are now implemented as:
\begin{subequations}\begin{align}
\MSE_u &= \frac{1}{N}\sum_{i = 1}^{N} |u^* - u_\mathrm{train}^*|^2 , \\
\MSE_{\phi}  &= \frac{1}{N}\sum_{i = 1}^{N} |\phi - \phi_\mathrm{train}|^2, \\
\MSE_m &= \frac{1}{N}\sum_{i = 1}^{N}\left|u^* - 1 + y^{*2} \right|^2,\\
\MSE_p &= \frac{1}{N}\sum_{i = 1}^{N}\left|K_c \phi \left(\phi \frac{d^2u^*}{dy^{*2}} + \frac{du^*}{dy^*} \frac{d\phi}{dy^*} \right) +  K_{\eta} \frac{du^*}{dy^*}\frac{\phi^2}{\eta_r}\frac{d\eta_r}{dy^*} + \frac{1}{Pe}\frac{d\phi}{dy^*} \right|^2, \label{eq:MSEs_Brownian} \\
\MSE_c  &= \left|\int_0^1 \phi(y^*) \,dy^* - \phi_b\right|^2.
\end{align}\label{eq:MSE_Brownian}\end{subequations}
Appendix~\ref{sec:pre-proc} describes how the experimental data (digitized from \citep{FAWM03}) was pre-processed into training data.

Using $\eta_r$ from Eq.~\eqref{eq:vis_kang}, 
we can calculate ${d\eta_r}/{dy^*} = ({d\eta_r}/{d\phi})({d\phi}/{dy^*})$ in Eq.~\eqref{eq:MSEs_Brownian} by the chain rule. Now, the loss $\mathscr{L}$ (recall Eq.~\eqref{eq:loss}) depends only on the neural networks' weights ($\bm{\Theta}_u$ and $\bm{\Theta}_\phi$) and the unknown model parameters $K_c$, $K_\eta$, and $K_{Pe}$. By minimizing the resulting $\mathscr{L}(\bm{\Theta}_u, \bm{\Theta}_\phi, K_c, K_\eta, K_{Pe})$ using $w_u = w_\phi = w_m = w_p = w_c = 1$,  with respect to its arguments, we find suitable neural networks' weights and values of the parameters $K_c$, $K_\eta$, and $K_{Pe}$ that simultaneously lead to the best agreement with the training data \emph{and} the physics.

\subsubsection{Comparison between PINN, theory and experiment}
For Brownian suspensions, we substitute the parabolic velocity profile from Eq.~\eqref{eq:momentum_poiseuille_parabolic} and the Brownian suspension viscosity from Eq.~\eqref{eq:vis_kang} into Eq.~\eqref{eq:pois_susp2}, to again obtain a nonlinear first-order ODE for $\phi(y^*)$:
\begin{equation}
    \frac{d \phi}{d y^*} = \frac{2K_c\phi^2 \eta_r +2y^*K_{\eta}\phi^2f_2(\phi)}{(Pe^{-1} -2K_cy^*\phi)\eta_r -2y^* K_{\eta} \phi^2 f_1(\phi)},
    \label{eq:ode_brownian}
\end{equation}
where 
\begin{subequations}\begin{align}
    f_1(\phi) &=\frac{d\eta_{\infty}}{d\phi} + \frac{{d\eta_0}/{d\phi} - {d\eta_{\infty}}/{d\phi}}{1 + K_{Pe} Pe\,y^*/(3\pi)}, \nonumber\\
    &=\frac{\mathfrak{a}_0}{\phi_{m_0}}\left(1-\frac{\phi}{\phi_{m_0}}\right)^{-\mathfrak{a}_0-1} + \frac{({\mathfrak{a}_0}/{\phi_{m_0}})(1-\phi/\phi_{m_0})^{-\mathfrak{a}_0-1} - ({\mathfrak{a}_\infty}/{\phi_{m_\infty}})(1-\phi/\phi_{m_\infty})^{-\mathfrak{a}_\infty-1} }{1 + K_{Pe} Pe\,y^*/(3\pi)},\\
    f_2(\phi) &= -[\eta_0(\phi) - \eta_{\infty}(\phi)]\frac{K_{Pe} Pe/(3\pi)}{[1 + K_{Pe} Pe\,y^*/(3\pi)]^2}.
\end{align}\label{eq:ode_brownian_supplement}\end{subequations}
Via the numerical procedure described in Sec.~\ref{sec:comparison_koh}, we solve for the ``theory'' prediction for $\phi(y^*)$ from Eqs.~\eqref{eq:ode_brownian} and \eqref{eq:phi_b_integral}, using the values for $K_c$ and $K_\eta$ in Eq.~\eqref{eq:ode_brownian} and $K_{Pe}$ in Eq.~\eqref{eq:ode_brownian_supplement} obtained by the PINN.

\begin{figure}[ht]
  \centering
  \includegraphics[width=\textwidth]{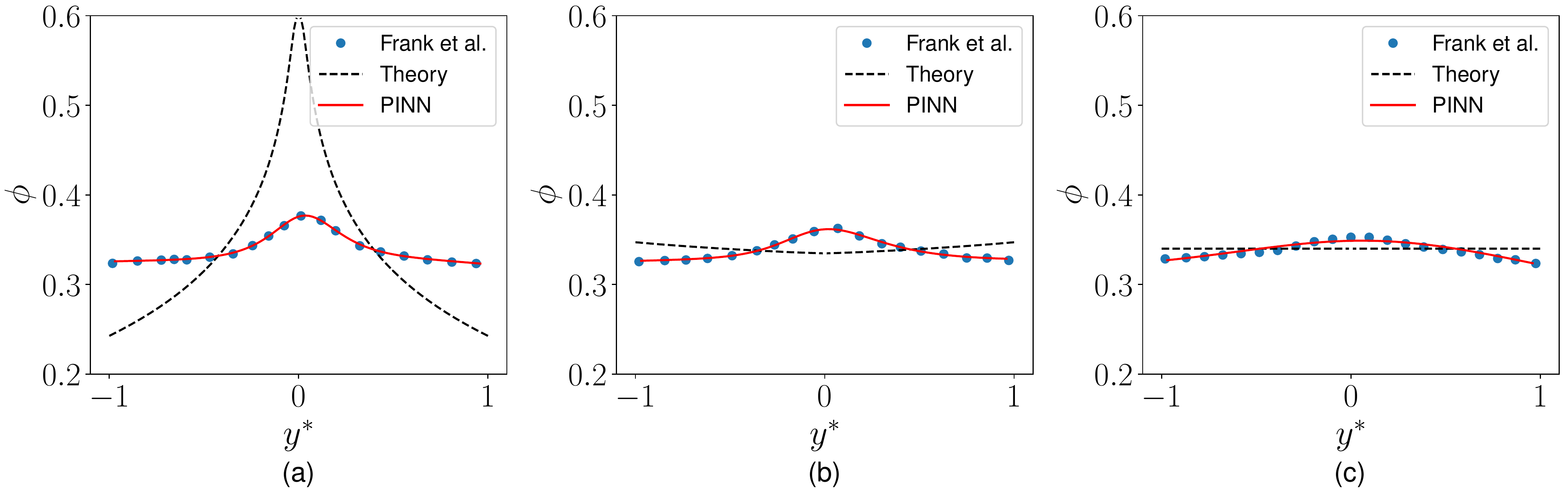}
\caption{Application to experiment from \citet{FAWM03} at $\phi_b = 0.34$: (a) $Pe = 4400$; (b) $Pe = 550$; (c) $Pe = 69$.}
\label{fig:frank_comp_34}
\end{figure}

\begin{figure}[ht]
  \centering
  \includegraphics[width=\textwidth]{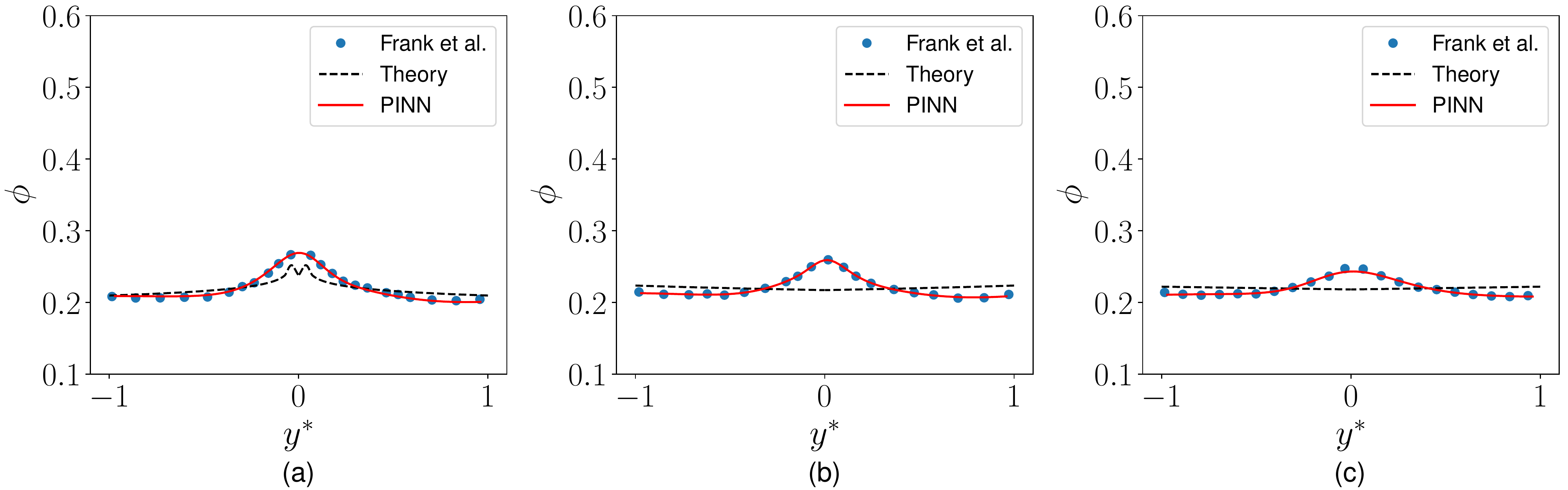}
\caption{Application to experiment from \citet{FAWM03} at $\phi_b = 0.22$: (a) $Pe = 4400$; (b) $Pe = 550$; (c) $Pe = 69$.}
\label{fig:frank_comp_22}
\end{figure}




\begin{table}[ht]
\centering
\begin{tabular}{ll|ccc}
\hline
\hline
$\phi_b$ & $Pe$ & $K_c$  & $K_\eta$  & $K_{Pe}$  \\ \hline
0.34     & $4400$ & $2.74 \times 10^{-4}$; $\in[0,5.64 \times 10^{-2}]$ & $8.28 \times 10^{-3}$; $\in[0,5.73 \times 10^{-1}]$  & $1.77$; $\in[1.70,2.52]$ \\ 
0.34     & $550$  & $7.84\times 10^{-5}$; $\in[0,8.94 \times 10^{-3}]$  & $6.32\times 10^{-4}$; $\in[0,1.09\times 10^{-1}]$    & $1.46$; $\in[1.16,1.94]$ \\ 
0.34     & $69$   & $2.47 \times 10^{-4}$; $\in[0,3.58 \times 10^{-3}]$   & $3.58 \times 10^{-7}$; $\in[0,7.09 \times 10^{-5}]$  & $1.51$; $\in[1.40,1.73]$ \\ \hline
0.22     & $4400$ & $4.17 \times 10^{-4}$; $\in[0,4.61 \times 10^{-2}]$  & $1.21 \times 10^{-2}$; $\in[0,7.84 \times 10^{-1}]$   & $1.71$; $\in[1.66,2.36]$ \\ 
0.22     & $550$  & $9.98 \times 10^{-5}$; $\in[0, 2.06 \times 10^{-2}]$   & $1.61 \times 10^{-3}$; $\in[0,2.37 \times 10^{-1}]$     & $ 1.47$; $\in[1.30,1.55]$\\   
0.22     & $69$  & $5.73 \times 10^{-4}$; $\in[0,9.85 \times 10^{-3}]$      & $5.51 \times 10^{-7}$; $\in[0,5.69 \times 10^{-6}]$     & $1.51$; $\in[1.44, 1.57]$ \\ \hline
\hline
\end{tabular}
\caption{Values of the shear-induced migration model's parameters $K_c$, $K_\eta$, and $K_{Pe}$, as inferred by the PINN from the experimental data of \citet{FAWM03}, for different bulk volume fractions $\phi_b$ and different values of shear P\'eclet number $Pe$. As before, these statistical results, with a mean and minimum/maximum interval, come from 1000 runs of the PINN algorithm using different random initializations of the NNs (excluding cases that failed to train).}
\label{tab:frank_params}
\end{table}

The comparisons between the experimental data of \citet{FAWM03} (symbols), the theory (dashed), and the PINN (solid) are shown in Figs.~\ref{fig:frank_comp_34} and \ref{fig:frank_comp_22}. The comparisons show that the PINN predictions agree well with experiment data, while the ``theory'' solutions do not. From discrete observations of $u^*$ and $\phi$ from the experiments used as training data, the PINN algorithm not only accurately predicts the distributions of $u^*(y^*)$ and $\phi(y^*)$, but also ``learns'' the suitable values of $K_c$, $K_\eta$, and $K_{Pe}$, which were \textit{a priori} unknown. 

Table~\ref{tab:frank_params} summarizes the values that the PINN ``learns'' from experimental data with different values of $\phi_b$ and $Pe$. To account for the variations in the converged values of the unknown model parameters, due to the random initialization of the neural networks, we averaged the predictions from 1000  initializations (excluding cases that failed to train, between $4\%$ and $6\%$ of the total, in which the loss was not minimized to the given tolerance) to obtain a statistical result with a mean and minimum/maximum interval. Importantly, the PINN analysis suggests that $K_c$ and $K_\eta$ vary with $\phi_b$ and $Pe$. This critical issue was \emph{not} addressed in previous works, in which the values of $K_c$ and $K_\eta$ are taken from \citep{pabga92} (validated only for the concentric Couette flow) and applied to any flow scenario. Now, however, we discover that $K_\eta$ (in particular) decreases with $Pe$. In addition, we observe that the model is quite sensitive when applied to Brownian suspensions, with the range of estimated values of $K_c$, $K_\eta$, and $K_{Pe}$ ranging from finite to close to zero. This observation may highlight that experiments measuring only the concentration profile may be insufficient to uniquely determine these several physical parameters (\textit{i.e.}, the problem is underdetermined).

Interestingly, the result for $Pe=69$ in Table~\ref{tab:frank_params} suggests that a vanishing viscosity-variation flux, $\|\bm{N}_\eta\|\approx0$, for this Brownian case of $Pe\not\gg1$. This result is consistent with the fact that the velocity profile is parabolic (recall Fig.~\ref{fig:frank_vel}), and the strongly-Brownian suspension effectively has the same viscosity as the Newtonian carrier fluid. Further, while the values obtained for $K_{Pe}$ in Table~\ref{tab:frank_params} are of the same order as the value $1.31$ used in literature, they are not the same, suggesting that this parameter (quantifying the suspension rheology's shear-dependence) should be measured for each  experiment, if possible.

Note that the model from Eq.~\eqref{eq:ode_brownian} breaks down if $d\phi/dy^*$ changes sign at some $y^*\ne0$. This situation can occur when the denominator in Eq.~\eqref{eq:ode_brownian} reaches $0$. The $\phi(y^*)$ profile develops a seemingly nonphysical maximum on each side of the centerline $y^*=0$ (see Fig.~\ref{fig:frank_comp_22}(a)). This observation highlights a deficiency in using the \citet{pabga92} model for Brownian suspensions. Further, it is important to emphasize that this breakdown of the Brownian shear-induced migration model is unrelated to the fact that the shear rate vanishes at the center of the channel, which is a separate issue addressed by ``nonlocal'' shear rate modifications that account for non-continuum effects expected to arise at the scale of a single particle diameter \cite{NB94,MB1999,MS95}. 
On the other hand, the PINN approach predicts a smooth curve that agrees with the experimental data because the PINN does not attempt to interpret the model as a ``basic law'' of fluid mechanics (which it is clearly not), but rather the PINN balances between fitting the experimental data and minimizing the model residual during the learning process.

\section{Conclusion}
\label{sec:conclusion}

Thirty years later, the phenomenological model of \citet{pabga92} continues to be the ``workhorse'' of macroscopic modeling of shear-induced particle migration in low-Reynolds-number flows of suspensions, as recent studies on simulation of migration in Brownian suspensions \citep{KM20} and experiments \citep{FSM20} and simulations \citep{H21} on migration in complex fluids demonstrate. However, the model's parameters were only ever properly calibrated against experiments in an annular Couette cell \citep{pabga92}. Subsequent studies on shear-induced migration in Poiseuille flow \citep{KHL1994,FAWM03} showed that the model, as calibrated against the annular Couette flow data, is only in \emph{qualitative} agreement with slot-flow experiments (despite being quantitatively accurate for Couette flow). 

To remedy these apparent contradictions/difficulties exposed in the literature, we proposed a new methodology in which the model of \citet{pabga92} is used to constrain a machine learning approach to assimilating the experimental data on particle migration. Using the idea of physics-informed neural networks (PINNs) pioneered by Karniadakis and collaborators \citep{RPK19,KKLPWY21}, we constructed a loss function from the  model of \citet{pabga92} and optimized neural networks to simultaneously best-approximate velocity and volume fraction experimental data, as well as the unknown/phenomenological parameters of the model. The PINN approach seamlessly identified the unknown parameters as part of the training process, extending the validity of the model of \citet{pabga92} to planar geometries and Brownian suspensions. In doing so, we found that the parameter values calibrated against Poiseuille flow and/or for Brownian suspension data differ from those calibrated against annular Couette flow experimental data. Additionally, the model's parameters were found to vary with the bulk volume fraction and the shear P\'eclet number of the suspension, which was not previously established for this model (though the P\'eclet dependence was observed in experiments \cite{FAWM03}, and in the suspension balance model discussed therein). This point was particularly important for the case of a Brownian suspension, highlighting why the phenomenological model solved as ``basic law'' with the parameters from \cite{pabga92} (as done in \cite{KM20}) could not match any of the experimental data.

In summary, we proposed to shift the paradigm of how phenomenological models for shear-induced migration should be used. The models widely used in the literature are only \emph{post}dictive, requiring calibration against an experiment for each flow scenario they are to be used in. Even then, attempting to solve the models as a ``basic law'' to predict the particle distribution (having somehow best fit the parameters) requires overcoming nonphysical singularities. On the other hand, employing the models within the PINN approach is not affected by mathematical singularities and does not need the parameters to be known \textit{a priori}. Therefore, their values and, thus, the relative importance of the different particle migration fluxes (collisional, viscosity-gradient, or Brownian) in a given experiment can be uncovered via PINNs (but not via the standard approach in the literature, based on directly solving an ODE for the particle distribution). It should  be re-emphasized that using the parameter values (calibrated in 1992 only for annular Couette flow) in varied flow scenarios strongly enforces physics that \emph{may or may not} be manifested in the particular flow under consideration. We have demonstrated that to gain an understanding of the ``unknown physics'' (to use the terminology of \citet{RHPT20}) of particle migration in a variety of flow experiments, PINNs can be effectively employed to simultaneously solve the inverse and forward problems and to significantly extend the practical utility of the standard phenomenological models.

\section*{Data accessibility statement}
The data and codes that support and reproduce the findings presented in this article are openly available in the repository at \url{http://doi.org/10.5281/zenodo.5735498} \cite{pinnRepo}.

\section*{Acknowledgments}
Acknowledgment is made to the donors of the American Chemical Society Petroleum Research Fund for support of the initial stages of this research under ACS PRF award \# 57371-DNI9.

\bibliography{references_ICC.bib}

\begin{thebibliography}{53}
\providecommand{\natexlab}[1]{#1}
\providecommand{\url}[1]{\texttt{#1}}
\providecommand{\urlprefix}{URL }
\expandafter\ifx\csname urlstyle\endcsname\relax
  \providecommand{\doi}[1]{doi:\discretionary{}{}{}#1}\else
  \providecommand{\doi}[1]{doi:\discretionary{}{}{}\begingroup
  \urlstyle{rm}\url{#1}\endgroup}\fi
\providecommand{\bibinfo}[2]{#2}

\bibitem[{Abadi et~al.(2015)Abadi, Agarwal, Barham, Brevdo, Chen, Citro,
  Corrado, Davis, Dean, Devin, Ghemawat, Goodfellow, Harp, Irving, Isard, Jia,
  Jozefowicz, Kaiser, Kudlur, Levenberg, Man{\'{e}}, Monga, Moore, Murray,
  Olah, Schuster, Shlens, Steiner, Sutskever, Talwar, Tucker, Vanhoucke,
  Vasudevan, Vi{\'{e}}gas, Vinyals, Warden, Wattenberg, Wicke, Yu, and
  Zheng}]{tensorflow2015-whitepaper}
\bibinfo{author}{M.~Abadi}, \bibinfo{author}{A.~Agarwal},
  \bibinfo{author}{P.~Barham}, \bibinfo{author}{E.~Brevdo},
  \bibinfo{author}{Z.~Chen}, \bibinfo{author}{C.~Citro}, \bibinfo{author}{G.~S.
  Corrado}, \bibinfo{author}{A.~Davis}, \bibinfo{author}{J.~Dean},
  \bibinfo{author}{M.~Devin}, \bibinfo{author}{S.~Ghemawat},
  \bibinfo{author}{I.~Goodfellow}, \bibinfo{author}{A.~Harp},
  \bibinfo{author}{G.~Irving}, \bibinfo{author}{M.~Isard},
  \bibinfo{author}{Y.~Jia}, \bibinfo{author}{R.~Jozefowicz},
  \bibinfo{author}{L.~Kaiser}, \bibinfo{author}{M.~Kudlur},
  \bibinfo{author}{K.~Levenberg}, \bibinfo{author}{D.~Man{\'{e}}},
  \bibinfo{author}{R.~Monga}, \bibinfo{author}{S.~Moore},
  \bibinfo{author}{D.~Murray}, \bibinfo{author}{C.~Olah},
  \bibinfo{author}{M.~Schuster}, \bibinfo{author}{J.~Shlens},
  \bibinfo{author}{B.~Steiner}, \bibinfo{author}{I.~Sutskever},
  \bibinfo{author}{K.~Talwar}, \bibinfo{author}{P.~Tucker},
  \bibinfo{author}{V.~Vanhoucke}, \bibinfo{author}{V.~Vasudevan},
  \bibinfo{author}{F.~Vi{\'{e}}gas}, \bibinfo{author}{O.~Vinyals},
  \bibinfo{author}{P.~Warden}, \bibinfo{author}{M.~Wattenberg},
  \bibinfo{author}{M.~Wicke}, \bibinfo{author}{Y.~Yu},
  \bibinfo{author}{X.~Zheng}, \bibinfo{title}{{TensorFlow: Large-Scale Machine
  Learning on Heterogeneous Systems}},
  \urlprefix\url{https://www.tensorflow.org/}, \bibinfo{year}{2015}.

\bibitem[{Alwosheel et~al.(2018)Alwosheel, van Cranenburgh, and Chorus}]{acc18}
\bibinfo{author}{A.~Alwosheel}, \bibinfo{author}{S.~van Cranenburgh},
  \bibinfo{author}{C.~G. Chorus}, \bibinfo{title}{{Is your dataset big enough?
  Sample size requirements when using artificial neural networks for discrete
  choice analysis}}, \bibinfo{journal}{J. Choice Model.} \bibinfo{volume}{28}
  (\bibinfo{year}{2018}) \bibinfo{pages}{167--182},
  \doi{\bibinfo{doi}{10.1016/j.jocm.2018.07.002}}.

\bibitem[{Baydin et~al.(2018)Baydin, Pearlmutter, Radul, and Siskind}]{BPRS18}
\bibinfo{author}{A.~G. Baydin}, \bibinfo{author}{B.~A. Pearlmutter},
  \bibinfo{author}{A.~A. Radul}, \bibinfo{author}{J.~M. Siskind},
  \bibinfo{title}{{Automatic Differentiation in Machine Learning: a Survey}},
  \bibinfo{journal}{J. Mach. Learning Res.}
  \bibinfo{volume}{18}~(\bibinfo{number}{153}) (\bibinfo{year}{2018})
  \bibinfo{pages}{1--43},
  \urlprefix\url{http://jmlr.org/papers/v18/17-468.html}.

\bibitem[{Bricker and Butler(2006)}]{BB06}
\bibinfo{author}{J.~M. Bricker}, \bibinfo{author}{J.~E. Butler},
  \bibinfo{title}{{Oscillatory shear of suspensions of noncolloidal
  particles}}, \bibinfo{journal}{J. Rheol.} \bibinfo{volume}{50}
  (\bibinfo{year}{2006}) \bibinfo{pages}{711--728},
  \doi{\bibinfo{doi}{10.1122/1.2234366}}.

\bibitem[{Brunton and Kutz(2019)}]{BK19_book}
\bibinfo{author}{S.~L. Brunton}, \bibinfo{author}{J.~N. Kutz},
  \bibinfo{title}{{Data-Driven Science and Engineering}},
  \bibinfo{publisher}{Cambridge University Press}, \bibinfo{address}{Cambridge,
  UK}, \doi{\bibinfo{doi}{10.1017/9781108380690}}, \bibinfo{year}{2019}.

\bibitem[{Brunton et~al.(2020)Brunton, Noack, and Koumoutsakos}]{BNK20}
\bibinfo{author}{S.~L. Brunton}, \bibinfo{author}{B.~R. Noack},
  \bibinfo{author}{P.~Koumoutsakos}, \bibinfo{title}{{Machine Learning for
  Fluid Mechanics}}, \bibinfo{journal}{Annu. Rev. Fluid Mech.}
  \bibinfo{volume}{52} (\bibinfo{year}{2020}) \bibinfo{pages}{477--508},
  \doi{\bibinfo{doi}{10.1146/annurev-fluid-010719-060214}}.

\bibitem[{Cai et~al.(2021)Cai, Wang, Wang, Perdikaris, and
  Karniadakis}]{CWWPK21}
\bibinfo{author}{S.~Cai}, \bibinfo{author}{Z.~Wang}, \bibinfo{author}{S.~Wang},
  \bibinfo{author}{P.~Perdikaris}, \bibinfo{author}{G.~E. Karniadakis},
  \bibinfo{title}{{Physics-Informed Neural Networks for Heat Transfer
  Problems}}, \bibinfo{journal}{ASME J. Heat Transfer} \bibinfo{volume}{143}
  (\bibinfo{year}{2021}) \bibinfo{pages}{060801},
  \doi{\bibinfo{doi}{10.1115/1.4050542}}.

\bibitem[{de~Kruif et~al.(1985)de~Kruif, van Iersel, Vrij, and Russel}]{KIV85}
\bibinfo{author}{C.~G. de~Kruif}, \bibinfo{author}{E.~M.~F. van Iersel},
  \bibinfo{author}{A.~Vrij}, \bibinfo{author}{W.~B. Russel},
  \bibinfo{title}{{Hard sphere colloidal dispersions: Viscosity as a function
  of shear rate and volume fraction}}, \bibinfo{journal}{J. Chem. Phys.}
  \bibinfo{volume}{83} (\bibinfo{year}{1985}) \bibinfo{pages}{4717--4725},
  \doi{\bibinfo{doi}{10.1063/1.448997}}.

\bibitem[{Denn and Morris(2014)}]{DM14}
\bibinfo{author}{M.~M. Denn}, \bibinfo{author}{J.~F. Morris},
  \bibinfo{title}{{Rheology of non-Brownian suspensions}},
  \bibinfo{journal}{Ann. Rev. Chem. Biomolec. Eng.} \bibinfo{volume}{5}
  (\bibinfo{year}{2014}) \bibinfo{pages}{203--228},
  \doi{\bibinfo{doi}{10.1146/annurev-chembioeng-060713-040221}}.

\bibitem[{Fang et~al.(2002)Fang, Mammoli, Brady, Ingber, Mondy, and
  Graham}]{FMBIMG02}
\bibinfo{author}{Z.~Fang}, \bibinfo{author}{A.~A. Mammoli},
  \bibinfo{author}{J.~F. Brady}, \bibinfo{author}{M.~S. Ingber},
  \bibinfo{author}{L.~A. Mondy}, \bibinfo{author}{A.~L. Graham},
  \bibinfo{title}{{Flow-aligned tensor models for suspension flows}},
  \bibinfo{journal}{Int. J. Multiphase Flow} \bibinfo{volume}{28}
  (\bibinfo{year}{2002}) \bibinfo{pages}{137--166},
  \doi{\bibinfo{doi}{10.1016/S0301-9322(01)00055-6}}.

\bibitem[{Fataei et~al.(2020)Fataei, Secrieru, and Mechtcherine}]{FSM20}
\bibinfo{author}{S.~Fataei}, \bibinfo{author}{E.~Secrieru},
  \bibinfo{author}{V.~Mechtcherine}, \bibinfo{title}{{Experimental Insights
  into Concrete Flow-Regimes Subject to Shear-Induced Particle Migration (SIPM)
  during Pumping}}, \bibinfo{journal}{Materials} \bibinfo{volume}{13}
  (\bibinfo{year}{2020}) \bibinfo{pages}{1233},
  \doi{\bibinfo{doi}{10.3390/ma13051233}}.

\bibitem[{Frank et~al.(2003)Frank, Anderson, Weeks, and Morris}]{FAWM03}
\bibinfo{author}{M.~Frank}, \bibinfo{author}{D.~Anderson},
  \bibinfo{author}{E.~R. Weeks}, \bibinfo{author}{J.~Morris},
  \bibinfo{title}{{Particle migration in pressure-driven flow of a Brownian
  suspension}}, \bibinfo{journal}{J. Fluid Mech.} \bibinfo{volume}{493}
  (\bibinfo{year}{2003}) \bibinfo{pages}{363--378},
  \doi{\bibinfo{doi}{10.1017/S0022112003006001}}.

\bibitem[{Gadala-Maria and Acrivos(1980)}]{GMA1980}
\bibinfo{author}{F.~Gadala-Maria}, \bibinfo{author}{A.~Acrivos},
  \bibinfo{title}{{Shear-induced structure in a concentrated suspension of
  solid spheres}}, \bibinfo{journal}{J. Rheol.} \bibinfo{volume}{24}
  (\bibinfo{year}{1980}) \bibinfo{pages}{799--814},
  \doi{\bibinfo{doi}{10.1122/1.549584}}.

\bibitem[{Guazzelli and Morris(2011)}]{GuaMor}
\bibinfo{author}{E.~Guazzelli}, \bibinfo{author}{J.~F. Morris},
  \bibinfo{title}{{A Physical Introduction to Suspension Dynamics}},
  vol.~\bibinfo{volume}{45} of \emph{\bibinfo{series}{Cambridge Texts in
  Applied Mathematics}}, \bibinfo{publisher}{Cambridge University Press},
  \bibinfo{address}{New York}, \doi{\bibinfo{doi}{10.1017/CBO9780511894671}},
  \bibinfo{year}{2011}.

\bibitem[{Haghighat et~al.(2021)Haghighat, Raissi, Moure, Gomez, and
  Juanes}]{HRMGJ21}
\bibinfo{author}{E.~Haghighat}, \bibinfo{author}{M.~Raissi},
  \bibinfo{author}{A.~Moure}, \bibinfo{author}{H.~Gomez},
  \bibinfo{author}{R.~Juanes}, \bibinfo{title}{{A physics-informed deep
  learning framework for inversion and surrogate modeling in solid mechanics}},
  \bibinfo{journal}{Comput. Meth. Appl. Mech. Engng} \bibinfo{volume}{379}
  (\bibinfo{year}{2021}) \bibinfo{pages}{113741},
  \doi{\bibinfo{doi}{10.1016/j.cma.2021.113741}}.

\bibitem[{He and Tartakovsky(2021)}]{HT20}
\bibinfo{author}{Q.~He}, \bibinfo{author}{A.~M. Tartakovsky},
  \bibinfo{title}{{Physics‐Informed Neural Network Method for Forward and
  Backward Advection‐Dispersion Equations}}, \bibinfo{journal}{Water Res.
  Res.} \bibinfo{volume}{57} (\bibinfo{year}{2021})
  \bibinfo{pages}{e2020WR029479}, \doi{\bibinfo{doi}{10.1029/2020WR029479}}.

\bibitem[{Henkes et~al.(2022)Henkes, Wessels, and Mahnken}]{HWM22}
\bibinfo{author}{A.~Henkes}, \bibinfo{author}{H.~Wessels},
  \bibinfo{author}{R.~Mahnken}, \bibinfo{title}{{Physics informed neural
  networks for continuum micromechanics}}, \bibinfo{journal}{Comput. Meth.
  Appl. Mech. Engng} \bibinfo{volume}{393} (\bibinfo{year}{2022})
  \bibinfo{pages}{114790}, \doi{\bibinfo{doi}{10.1016/j.cma.2022.114790}}.

\bibitem[{Hern{\'{a}}ndez(2021)}]{H21}
\bibinfo{author}{R.~Hern{\'{a}}ndez}, \bibinfo{title}{{Dynamics of concentrated
  suspensions in two-dimensional channel flow for non-Newtonian slurries}},
  \bibinfo{journal}{Int. J. Multiphase Flow} \bibinfo{volume}{139}
  (\bibinfo{year}{2021}) \bibinfo{pages}{103616},
  \doi{\bibinfo{doi}{10.1016/j.ijmultiphaseflow.2021.103616}}.

\bibitem[{Jana et~al.(1995)Jana, Kapoor, and Acrivos}]{JKA95}
\bibinfo{author}{S.~C. Jana}, \bibinfo{author}{B.~Kapoor},
  \bibinfo{author}{A.~Acrivos}, \bibinfo{title}{{Apparent wall slip velocity
  coefficients in concentrated suspensions of noncolloidal particles}},
  \bibinfo{journal}{J. Rheol.} \bibinfo{volume}{39} (\bibinfo{year}{1995})
  \bibinfo{pages}{1123--1132}, \doi{\bibinfo{doi}{10.1122/1.550631}}.

\bibitem[{Jin et~al.(2021)Jin, Cai, Li, and Karniadakis}]{JCLK20}
\bibinfo{author}{X.~Jin}, \bibinfo{author}{S.~Cai}, \bibinfo{author}{H.~Li},
  \bibinfo{author}{G.~E. Karniadakis}, \bibinfo{title}{{NSFnets (Navier-Stokes
  flow nets): Physics-informed neural networks for the incompressible
  Navier-Stokes equations}}, \bibinfo{journal}{J. Comput. Phys.}
  \bibinfo{volume}{426} (\bibinfo{year}{2021}) \bibinfo{pages}{109951},
  \doi{\bibinfo{doi}{10.1016/j.jcp.2020.109951}}.

\bibitem[{Kalyon(2005)}]{K05}
\bibinfo{author}{D.~M. Kalyon}, \bibinfo{title}{{Apparent slip and
  viscoplasticity of concentrated suspensions}}, \bibinfo{journal}{J. Rheol.}
  \bibinfo{volume}{49} (\bibinfo{year}{2005}) \bibinfo{pages}{621--640},
  \doi{\bibinfo{doi}{10.1122/1.1879043}}.

\bibitem[{Kang and Mirbod(2020)}]{KM20}
\bibinfo{author}{C.~Kang}, \bibinfo{author}{P.~Mirbod},
  \bibinfo{title}{{Shear-induced particle migration of semi-dilute and
  concentrated Brownian suspensions in both Poiseuille and circular Couette
  flow}}, \bibinfo{journal}{Int. J. Multiphase Flow} \bibinfo{volume}{126}
  (\bibinfo{year}{2020}) \bibinfo{pages}{103239},
  \doi{\bibinfo{doi}{10.1016/j.ijmultiphaseflow.2020.103239}}.

\bibitem[{Karniadakis et~al.(2021)Karniadakis, Kevrekidis, Lu, Perdikaris,
  Wang, and Yang}]{KKLPWY21}
\bibinfo{author}{G.~E. Karniadakis}, \bibinfo{author}{I.~G. Kevrekidis},
  \bibinfo{author}{L.~Lu}, \bibinfo{author}{P.~Perdikaris},
  \bibinfo{author}{S.~Wang}, \bibinfo{author}{L.~Yang},
  \bibinfo{title}{{Physics-informed machine learning}}, \bibinfo{journal}{Nat.
  Rev. Phys.} \bibinfo{volume}{3} (\bibinfo{year}{2021})
  \bibinfo{pages}{422--440}, \doi{\bibinfo{doi}{10.1038/s42254-021-00314-5}}.

\bibitem[{Kingma and Ba(2015)}]{KB15}
\bibinfo{author}{D.~P. Kingma}, \bibinfo{author}{J.~Ba}, \bibinfo{title}{{Adam:
  A Method for Stochastic Optimization}}, in: \bibinfo{editor}{Y.~Bengio},
  \bibinfo{editor}{Y.~LeCun} (Eds.), \bibinfo{booktitle}{3rd International
  Conference on Learning Representations (ICLR)}, \bibinfo{address}{San Diego,
  CA}, \urlprefix\url{http://arxiv.org/abs/1412.6980}, \bibinfo{year}{2015}.

\bibitem[{Koh et~al.(1994)Koh, Hookham, and Leal}]{KHL1994}
\bibinfo{author}{C.~J. Koh}, \bibinfo{author}{P.~Hookham},
  \bibinfo{author}{L.~G. Leal}, \bibinfo{title}{{An experimental investigation
  of concentrated suspension flows in a rectangular channel}},
  \bibinfo{journal}{J. Fluid Mech.} \bibinfo{volume}{266}
  (\bibinfo{year}{1994}) \bibinfo{pages}{1--32},
  \doi{\bibinfo{doi}{10.1017/S0022112094000911}}.

\bibitem[{Krieger(1972)}]{Krieger72}
\bibinfo{author}{I.~M. Krieger}, \bibinfo{title}{{Rheology of monodisperse
  latices}}, \bibinfo{journal}{Adv. Colloid Interface Sci.} \bibinfo{volume}{3}
  (\bibinfo{year}{1972}) \bibinfo{pages}{111--136},
  \doi{\bibinfo{doi}{10.1016/0001-8686(72)80001-0}}.

\bibitem[{Krieger and Dougherty(1959)}]{KD59}
\bibinfo{author}{I.~M. Krieger}, \bibinfo{author}{T.~J. Dougherty},
  \bibinfo{title}{{A Mechanism for Non‐Newtonian Flow in Suspensions of Rigid
  Spheres}}, \bibinfo{journal}{Trans. Soc. Rheol.} \bibinfo{volume}{3}
  (\bibinfo{year}{1959}) \bibinfo{pages}{137--152},
  \doi{\bibinfo{doi}{10.1122/1.548848}}.

\bibitem[{Leighton and Acrivos(1987)}]{la87}
\bibinfo{author}{D.~Leighton}, \bibinfo{author}{A.~Acrivos},
  \bibinfo{title}{{The shear-induced migration of particles in concentrated
  suspensions}}, \bibinfo{journal}{J. Fluid Mech.} \bibinfo{volume}{181}
  (\bibinfo{year}{1987}) \bibinfo{pages}{415--439},
  \doi{\bibinfo{doi}{10.1017/S0022112087002155}}.

\bibitem[{Lhuillier(2009)}]{L09}
\bibinfo{author}{D.~Lhuillier}, \bibinfo{title}{{Migration of rigid particles
  in non-Brownian viscous suspensions}}, \bibinfo{journal}{Phys. Fluids}
  \bibinfo{volume}{21} (\bibinfo{year}{2009}) \bibinfo{pages}{023302},
  \doi{\bibinfo{doi}{10.1063/1.3079672}}.

\bibitem[{Li et~al.(2021)Li, Bazant, and Zhu}]{LBZ21}
\bibinfo{author}{W.~Li}, \bibinfo{author}{M.~Z. Bazant},
  \bibinfo{author}{J.~Zhu}, \bibinfo{title}{{A physics-guided neural network
  framework for elastic plates: Comparison of governing equations-based and
  energy-based approaches}}, \bibinfo{journal}{Comput. Meth. Appl. Mech. Engng}
  \bibinfo{volume}{383} (\bibinfo{year}{2021}) \bibinfo{pages}{113933},
  \doi{\bibinfo{doi}{10.1016/j.cma.2021.113933}}.

\bibitem[{Lu and Christov(2021)}]{pinnRepo}
\bibinfo{author}{D.~Lu}, \bibinfo{author}{I.~C. Christov},
  \bibinfo{title}{{PINN{\_}shear{\_}migration{\_}of{\_}particles}},
  \doi{\bibinfo{doi}{10.5281/zenodo.5735498}}, \bibinfo{year}{2021}.

\bibitem[{Mao et~al.(2020)Mao, Jagtap, and Karniadakis}]{MJK19}
\bibinfo{author}{Z.~Mao}, \bibinfo{author}{A.~D. Jagtap},
  \bibinfo{author}{G.~E. Karniadakis}, \bibinfo{title}{{Physics-informed neural
  networks for high-speed flows}}, \bibinfo{journal}{Comput. Meth. Appl. Mech.
  Engng} \bibinfo{volume}{360} (\bibinfo{year}{2020}) \bibinfo{pages}{112789},
  \doi{\bibinfo{doi}{10.1016/j.cma.2019.112789}}.

\bibitem[{Maxey(2017)}]{M17}
\bibinfo{author}{M.~Maxey}, \bibinfo{title}{{Simulation methods for particulate
  flows and concentrated suspensions}}, \bibinfo{journal}{Annu. Rev. Fluid
  Mech.} \bibinfo{volume}{49} (\bibinfo{year}{2017}) \bibinfo{pages}{171--193},
  \doi{\bibinfo{doi}{10.1146/annurev-fluid-122414-034408}}.

\bibitem[{Meng and Karniadakis(2020)}]{MK20}
\bibinfo{author}{X.~Meng}, \bibinfo{author}{G.~E. Karniadakis},
  \bibinfo{title}{{A composite neural network that learns from multi-fidelity
  data: Application to function approximation and inverse PDE problems}},
  \bibinfo{journal}{J. Comput. Phys.} \bibinfo{volume}{401}
  (\bibinfo{year}{2020}) \bibinfo{pages}{109020},
  \doi{\bibinfo{doi}{10.1016/j.jcp.2019.109020}}.

\bibitem[{Merhi et~al.(2005)Merhi, Lemaire, Bossis, and Moukalled}]{MLBM05}
\bibinfo{author}{D.~Merhi}, \bibinfo{author}{E.~Lemaire},
  \bibinfo{author}{G.~Bossis}, \bibinfo{author}{F.~Moukalled},
  \bibinfo{title}{{Particle migration in a concentrated suspension flowing
  between rotating parallel plates: Investigation of diffusion flux
  coefficients}}, \bibinfo{journal}{J. Rheol.} \bibinfo{volume}{49}
  (\bibinfo{year}{2005}) \bibinfo{pages}{1429--1448},
  \doi{\bibinfo{doi}{10.1122/1.2079247}}.

\bibitem[{Miller and Morris(2006)}]{MM2006}
\bibinfo{author}{R.~M. Miller}, \bibinfo{author}{J.~F. Morris},
  \bibinfo{title}{{Normal stress-driven migration and axial development in
  pressure-driven flow of concentrated suspensions}}, \bibinfo{journal}{J.
  Non-Newtonian Fluid Mech.} \bibinfo{volume}{135} (\bibinfo{year}{2006})
  \bibinfo{pages}{149--165}, \doi{\bibinfo{doi}{10.1016/J.JNNFM.2005.11.009}}.

\bibitem[{Mills and Snabre(1995)}]{MS95}
\bibinfo{author}{P.~Mills}, \bibinfo{author}{P.~Snabre},
  \bibinfo{title}{{Rheology and structure of concentrated suspensions of hard
  spheres. Shear induced particle migration}}, \bibinfo{journal}{J. Physique
  II} \bibinfo{volume}{5} (\bibinfo{year}{1995}) \bibinfo{pages}{1597--1608},
  \doi{\bibinfo{doi}{10.1051/jp2:1995201}}.

\bibitem[{Monsorno et~al.(2017)Monsorno, Varsakelis, and Papalexandris}]{MVP17}
\bibinfo{author}{D.~Monsorno}, \bibinfo{author}{C.~Varsakelis},
  \bibinfo{author}{M.~Papalexandris}, \bibinfo{title}{{Poiseuille flow of dense
  non-colloidal suspensions: The role of intergranular and nonlocal stresses in
  particle migration}}, \bibinfo{journal}{J. Non-Newtonian Fluid Mech.}
  \bibinfo{volume}{247} (\bibinfo{year}{2017}) \bibinfo{pages}{229--238},
  \doi{\bibinfo{doi}{10.1016/j.jnnfm.2017.08.002}}.

\bibitem[{Morris(2020)}]{M20}
\bibinfo{author}{J.~F. Morris}, \bibinfo{title}{{Toward a fluid mechanics of
  suspensions}}, \bibinfo{journal}{Phys. Rev. Fluids} \bibinfo{volume}{5}
  (\bibinfo{year}{2020}) \bibinfo{pages}{110519},
  \doi{\bibinfo{doi}{10.1103/PhysRevFluids.5.110519}}.

\bibitem[{Morris and Boulay(1999)}]{MB1999}
\bibinfo{author}{J.~F. Morris}, \bibinfo{author}{F.~Boulay},
  \bibinfo{title}{{Curvilinear flows of noncolloidal suspensions: The role of
  normal stresses}}, \bibinfo{journal}{J. Rheol.} \bibinfo{volume}{43}
  (\bibinfo{year}{1999}) \bibinfo{pages}{1213--1237},
  \doi{\bibinfo{doi}{10.1122/1.551021}}.

\bibitem[{Municchi et~al.(2019)Municchi, Nagrani, and Christov}]{MNC19}
\bibinfo{author}{F.~Municchi}, \bibinfo{author}{P.~P. Nagrani},
  \bibinfo{author}{I.~C. Christov}, \bibinfo{title}{{A two-fluid model for
  numerical simulation of shear-dominated suspension flows}},
  \bibinfo{journal}{Int. J. Multiphase Flow} \bibinfo{volume}{120}
  (\bibinfo{year}{2019}) \bibinfo{pages}{103079},
  \doi{\bibinfo{doi}{10.1016/j.ijmultiphaseflow.2019.07.015}}.

\bibitem[{Nott and Brady(1994)}]{NB94}
\bibinfo{author}{P.~R. Nott}, \bibinfo{author}{J.~F. Brady},
  \bibinfo{title}{{Pressure-driven flow of suspensions: simulation and
  theory}}, \bibinfo{journal}{J. Fluid Mech.} \bibinfo{volume}{275}
  (\bibinfo{year}{1994}) \bibinfo{pages}{157--199},
  \doi{\bibinfo{doi}{10.1017/S0022112094002326}}.

\bibitem[{Nott et~al.(2011)Nott, Guazzelli, and Pouliquen}]{Nott2011}
\bibinfo{author}{P.~R. Nott}, \bibinfo{author}{E.~Guazzelli},
  \bibinfo{author}{O.~Pouliquen}, \bibinfo{title}{{The suspension balance model
  revisited}}, \bibinfo{journal}{Phys. Fluids} \bibinfo{volume}{23}
  (\bibinfo{year}{2011}) \bibinfo{pages}{043304},
  \doi{\bibinfo{doi}{10.1063/1.3570921}}.

\bibitem[{Panton(2013)}]{Panton}
\bibinfo{author}{R.~L. Panton}, \bibinfo{title}{{Incompressible flow}},
  \bibinfo{publisher}{John Wiley {\&} Sons}, \bibinfo{address}{Hoboken, NJ},
  \bibinfo{edition}{4th} edn., \doi{\bibinfo{doi}{10.1002/9781118713075}},
  \bibinfo{year}{2013}.

\bibitem[{Phillips et~al.(1992)Phillips, Armstrong, Brown, Graham, and
  Abbott}]{pabga92}
\bibinfo{author}{R.~J. Phillips}, \bibinfo{author}{R.~C. Armstrong},
  \bibinfo{author}{R.~A. Brown}, \bibinfo{author}{A.~L. Graham},
  \bibinfo{author}{J.~R. Abbott}, \bibinfo{title}{{A constitutive equation for
  concentrated suspensions that accounts for shear-induced particle
  migration}}, \bibinfo{journal}{Phys. Fluids A} \bibinfo{volume}{4}
  (\bibinfo{year}{1992}) \bibinfo{pages}{30--40},
  \doi{\bibinfo{doi}{10.1063/1.858498}}.

\bibitem[{Raissi et~al.(2019)Raissi, Perdikaris, and Karniadakis}]{RPK19}
\bibinfo{author}{M.~Raissi}, \bibinfo{author}{P.~Perdikaris},
  \bibinfo{author}{G.~Karniadakis}, \bibinfo{title}{{Physics-informed neural
  networks: A deep learning framework for solving forward and inverse problems
  involving nonlinear partial differential equations}}, \bibinfo{journal}{J.
  Comput. Phys.} \bibinfo{volume}{378} (\bibinfo{year}{2019})
  \bibinfo{pages}{686--707}, \doi{\bibinfo{doi}{10.1016/j.jcp.2018.10.045}}.

\bibitem[{Raissi et~al.(2020)Raissi, Yazdani, and Karniadakis}]{MYK20}
\bibinfo{author}{M.~Raissi}, \bibinfo{author}{A.~Yazdani},
  \bibinfo{author}{G.~E. Karniadakis}, \bibinfo{title}{{Hidden fluid mechanics:
  Learning velocity and pressure fields from flow visualizations}},
  \bibinfo{journal}{Science} \bibinfo{volume}{367} (\bibinfo{year}{2020})
  \bibinfo{pages}{1026--1030}, \doi{\bibinfo{doi}{10.1126/science.aaw4741}}.

\bibitem[{Reyes et~al.(2021)Reyes, Howard, Perdikaris, and
  Tartakovsky}]{RHPT20}
\bibinfo{author}{B.~Reyes}, \bibinfo{author}{A.~A. Howard},
  \bibinfo{author}{P.~Perdikaris}, \bibinfo{author}{A.~M. Tartakovsky},
  \bibinfo{title}{{Learning unknown physics of non-Newtonian fluids}},
  \bibinfo{journal}{Phys. Rev. Fluids} \bibinfo{volume}{6}
  (\bibinfo{year}{2021}) \bibinfo{pages}{073301},
  \doi{\bibinfo{doi}{10.1103/PhysRevFluids.6.073301}}.

\bibitem[{Stickel and Powell(2005)}]{SP05}
\bibinfo{author}{J.~J. Stickel}, \bibinfo{author}{R.~L. Powell},
  \bibinfo{title}{{Fluid mechanics and rheology of dense suspensions}},
  \bibinfo{journal}{Ann. Rev. Fluid Mech.} \bibinfo{volume}{37}
  (\bibinfo{year}{2005}) \bibinfo{pages}{129--149},
  \doi{\bibinfo{doi}{10.1146/annurev.fluid.36.050802.122132}}.

\bibitem[{Virtanen et~al.(2020)Virtanen, Gommers, Oliphant, Haberland, Reddy,
  Cournapeau, Burovski, Peterson, Weckesser, Bright, van~der Walt, Brett,
  Wilson, Millman, Mayorov, Nelson, Jones, Kern, Larson, Carey, Polat, Feng,
  Moore, VanderPlas, Laxalde, Perktold, Cimrman, Henriksen, Quintero, Harris,
  Archibald, Ribeiro, Pedregosa, and van Mulbregt}]{SciPy}
\bibinfo{author}{P.~Virtanen}, \bibinfo{author}{R.~Gommers},
  \bibinfo{author}{T.~E. Oliphant}, \bibinfo{author}{M.~Haberland},
  \bibinfo{author}{T.~Reddy}, \bibinfo{author}{D.~Cournapeau},
  \bibinfo{author}{E.~Burovski}, \bibinfo{author}{P.~Peterson},
  \bibinfo{author}{W.~Weckesser}, \bibinfo{author}{J.~Bright},
  \bibinfo{author}{S.~J. van~der Walt}, \bibinfo{author}{M.~Brett},
  \bibinfo{author}{J.~Wilson}, \bibinfo{author}{K.~J. Millman},
  \bibinfo{author}{N.~Mayorov}, \bibinfo{author}{A.~R.~J. Nelson},
  \bibinfo{author}{E.~Jones}, \bibinfo{author}{R.~Kern},
  \bibinfo{author}{E.~Larson}, \bibinfo{author}{C.~J. Carey},
  \bibinfo{author}{I.~Polat}, \bibinfo{author}{Y.~Feng}, \bibinfo{author}{E.~W.
  Moore}, \bibinfo{author}{J.~VanderPlas}, \bibinfo{author}{D.~Laxalde},
  \bibinfo{author}{J.~Perktold}, \bibinfo{author}{R.~Cimrman},
  \bibinfo{author}{I.~Henriksen}, \bibinfo{author}{E.~A. Quintero},
  \bibinfo{author}{C.~R. Harris}, \bibinfo{author}{A.~M. Archibald},
  \bibinfo{author}{A.~H. Ribeiro}, \bibinfo{author}{F.~Pedregosa},
  \bibinfo{author}{P.~van Mulbregt}, \bibinfo{title}{{SciPy 1.0: fundamental
  algorithms for scientific computing in Python}}, \bibinfo{journal}{Nat.
  Methods} \bibinfo{volume}{17} (\bibinfo{year}{2020})
  \bibinfo{pages}{261--272}, \doi{\bibinfo{doi}{10.1038/s41592-019-0686-2}}.

\bibitem[{Vollebregt et~al.(2010)Vollebregt, Van Der~Sman, and Boom}]{vsb10}
\bibinfo{author}{H.~M. Vollebregt}, \bibinfo{author}{R.~G.~M. Van Der~Sman},
  \bibinfo{author}{R.~M. Boom}, \bibinfo{title}{{Suspension flow modelling in
  particle migration and microfiltration}}, \bibinfo{journal}{Soft Matter}
  \bibinfo{volume}{6} (\bibinfo{year}{2010}) \bibinfo{pages}{6052--6064},
  \doi{\bibinfo{doi}{10.1039/c0sm00217h}}.

\bibitem[{Wang et~al.(2022)Wang, Yu, and Perdikaris}]{WYP20}
\bibinfo{author}{S.~Wang}, \bibinfo{author}{X.~Yu},
  \bibinfo{author}{P.~Perdikaris}, \bibinfo{title}{{When and why PINNs fail to
  train: A neural tangent kernel perspective}}, \bibinfo{journal}{J. Comput.
  Phys.} \bibinfo{volume}{449} (\bibinfo{year}{2022}) \bibinfo{pages}{110768},
  \doi{\bibinfo{doi}{10.1016/j.jcp.2021.110768}}.

\bibitem[{Yang et~al.(2019)Yang, Zafar, Wang, and Xiao}]{YZWX19}
\bibinfo{author}{X.~I.~A. Yang}, \bibinfo{author}{S.~Zafar},
  \bibinfo{author}{J.-X. Wang}, \bibinfo{author}{H.~Xiao},
  \bibinfo{title}{{Predictive large-eddy-simulation wall modeling via
  physics-informed neural networks}}, \bibinfo{journal}{Phys. Rev. Fluids}
  \bibinfo{volume}{4} (\bibinfo{year}{2019}) \bibinfo{pages}{034602},
  \doi{\bibinfo{doi}{10.1103/PhysRevFluids.4.034602}}.

\end{thebibliography}

\clearpage

\appendix
\section*{Appendix}

\section{Pre-processing of experimental data into training data}
\label{sec:pre-proc}

For the training of the NNs, we utilize the experimental data from the literature. Specifically, we digitized the data from Fig.~7 of Ref.~\cite{pabga92}, Figs.~10, 11, 15, and 19 of Ref.~\cite{KHL1994}, and Figs.~3 and 4 of Ref.~\cite{FAWM03}. Experimental data points in these figures are limited (approximately 20 points in each plot). The NNs need more data points to achieve successful training (approximately 50 times the number of weights \cite{acc18}). Thus, assuming the particle migration profiles and suspension velocities are smooth functions, we use interpolation to generate $2000$ data samples for the training of the NNs from the limited experimental data points. Specifically, we use \texttt{interpolate.interp1d} from the SciPy stack in Python \citep{SciPy} to obtain values at sample points that are not part of the digitized experimental data.

\section{Choice of number of hidden layers and neurons per layer}
\label{sec:choice_layers}

\begin{figure}[ht]
  \centering
  \includegraphics[width=0.65\textwidth]{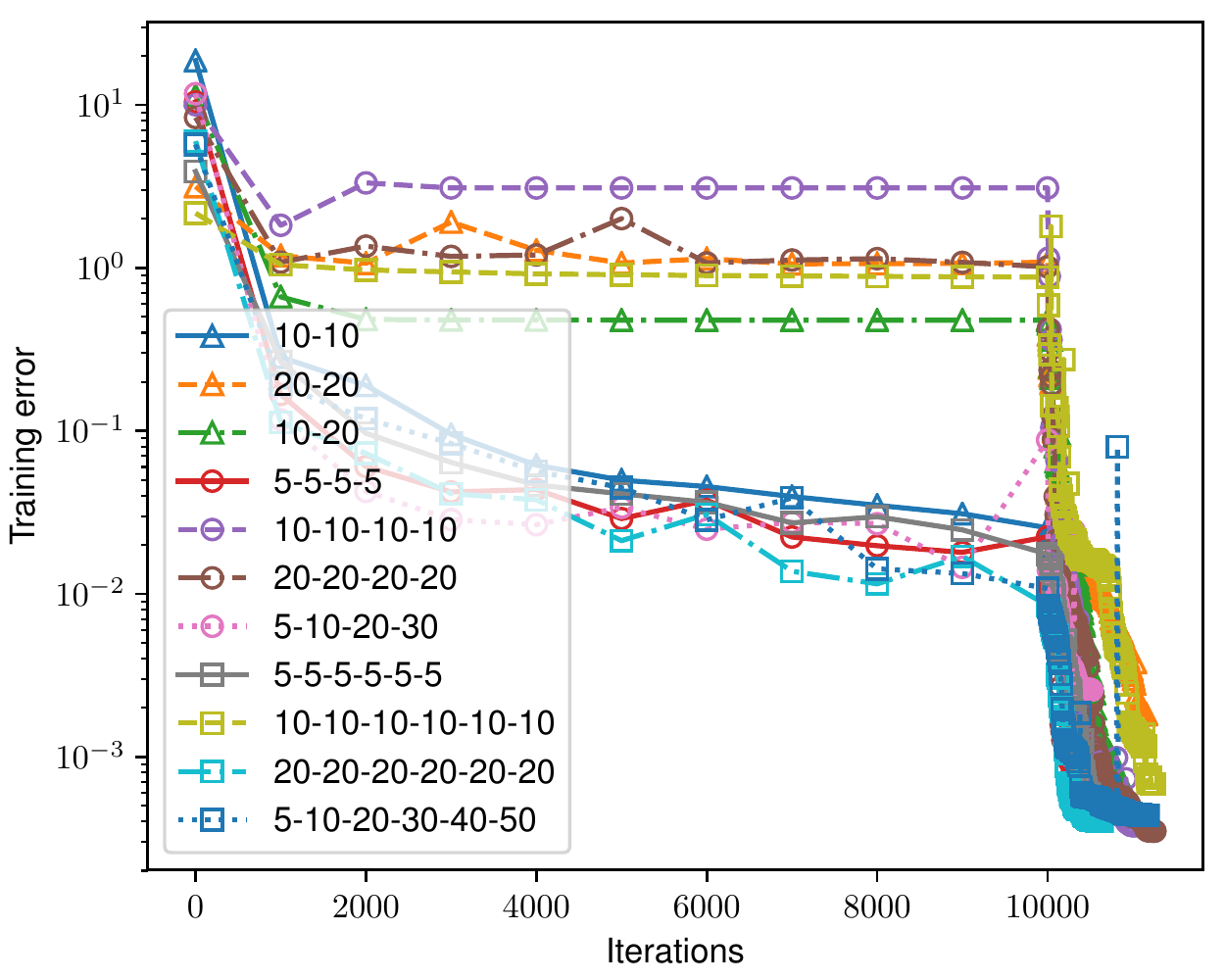}
  \caption{Training errors from PINNs with different numbers of hidden layers, and a different number of neurons in each layer, in the NN architecture. The notation in the plot labels represents the number of neurons per layer (and, thus, the total number of layers).}
\label{fig:couette_errors}
\end{figure}

To attempt to find an optimal number of hidden layers and nodes per layer (and justify the choices made for the PINN architecture used in the main text), we plot the training error for different numbers of hidden layers, as shown in Fig.~\ref{fig:couette_errors}. We first train the PINNs for $10,000$ iterations using the ``Adam'' \citep{KB15} optimizer, then we use the ``L-BFGS-B'' optimizer until convergence. Figure~\ref{fig:couette_errors} shows that adding layers (and/or more neurons per layer) does not reduce the training error further, while this action leads to a significantly higher computational cost (and requires more computing resources for the training process to reach convergence). Using fewer layers (and/or fewer neurons per layer) tends to lead to divergence, i.e., failure of the training process. Therefore, as a suitable trade-off, we use 2 hidden layers with 10 neurons in each layer in the NNs of the PINNs.

\end{document}